\documentclass{aa}  

\usepackage{booktabs}
\usepackage{graphicx}
\usepackage[english]{babel}
\usepackage{multirow}
\usepackage{xcolor}
\usepackage{natbib} 
\usepackage{gensymb}
\usepackage{ftnxtra}
\usepackage{fnpos}
\usepackage{txfonts}
\usepackage{soul}
\setstcolor{red}

\begin{document}

   \title{Analysis of apsidal motion in eclipsing binaries using {\em TESS} data\thanks{Table 3 is only available in electronic form at the CDS via anonymous ftp to cdsarc.u-strasbg.fr (130.79.128.5) or via http://cdsweb.u-strasbg.fr/cgi-bin/qcat?J/A+A/ }}

   \subtitle{I. A test of gravitational theories}

   \author{D. Baroch \inst{1,2}
          \and
          A. Gim\'enez \inst{3,4}
          \and
          I. Ribas \inst{1,2}
          \and
          J.~C. Morales \inst{1,2}
          \and
          G. Anglada-Escud\'e \inst{1,2}
          \and
          A.~Claret \inst{5}
          }

   \institute{
        Institut de Ci\`encies de l'Espai (ICE, CSIC),
        Campus UAB, c/ Can Magrans s/n, E-08193 Bellaterra, Barcelona, Spain, \email{baroch@ice.cat}
        \and
        Institut d'Estudis Espacials de Catalunya (IEEC),
        c/ Gran Capit\`a 2-4, E-08034 Barcelona, Spain
        \and
        Centro de Astrobiolog\'{\i}a (CSIC-INTA), E-28692 Villanueva de la Cañada, Madrid, Spain 
        \and
        International Space Science Institute (ISSI), Hallerstrasse 6, CH-3012 Bern, Switzerland
        \and
        Instituto de Astrof\'isica de Andaluc\'ia (IAA-CSIC), Glorieta de la Astronom\'ia s/n, E-18008 Granada, Spain 
             }

   \date{Received November 27, 2020; accepted March 2, 2021}

 
  \abstract
   {The change in the argument of periastron of eclipsing binaries, that is, the apsidal motion caused by classical and relativistic effects, can be measured from variations in the difference between the time of minimum light of the primary and secondary eclipses. Poor apsidal motion rate determinations and large uncertainties in the classical term have hampered previous attempts to determine the general relativistic term with sufficient precision to test general relativity predictions.}
   {As a product of the {\em TESS} mission, thousands of high-precision light curves from eclipsing binaries are now available. Using a selection of suitable well-studied eccentric eclipsing binary systems, we aim to determine their apsidal motion rates and place constraints on key gravitational parameters.}
   {We compute the time of minimum light from the {\em TESS} light curves of 15 eclipsing binaries with precise absolute parameters and with an expected general relativistic contribution to the total apsidal motion rate of greater than 60\%. We use the changing primary and secondary eclipse timing differences over time to compute the apsidal motion rate, when possible, or the difference between the linear periods as computed from primary and secondary eclipses. For a greater time baseline we carefully combine the high-precision {\em TESS} timings with archival reliable timings.}
   {We determine the apsidal motion rate of 9 eclipsing binaries, 5 of which are reported for the first time. From these, we are able to measure the general relativistic apsidal motion rate of 6 systems with sufficient precision to test general relativity for the first time using this method. This test explores a regime of gravitational forces and potentials that had not been probed before. We find perfect agreement with theoretical predictions, and we are able to set stringent constraints on two parameters of the parametrised post-Newtonian formalism.}
   {}


   \maketitle
%

\section{Introduction} \label{sec:intro}

Eclipsing binaries have demonstrated to be a basic source of fundamental information about stellar properties such as masses and radii \citep{Andersen1991,Torres2010}. The comparison of the observed values with theoretical predictions has been used
profusely to perform critical tests of stellar structure and evolution models \citep{Pols1997,Ribas2000,Torres2002,Lastennet2002,Feiden2012,Higl2017,Tkachenko2020}. But eccentric eclipsing binaries offer further opportunities to gain indirect insight into the internal structure of stars through the measurement of the precession rate of the line of apsides of the orbit, that is, the apsidal motion rate ($\dot{\omega}$). Such precession motion is found to arise from two different contributions: a general relativistic ($\dot{\omega}_{\rm rel}$) term arising from general relativity (GR), and a classical or Newtonian ($\dot{\omega}_{\rm cl}$) term. These two contributions are additive, so that $\dot{\omega}=\dot{\omega}_{\rm rel}+\dot{\omega}_{\rm cl}$ \citep{Shakura1985}. The general relativistic term of the apsidal motion rate, when only considering quadratic corrections, can be calculated with the equation given by \cite{Levi1937}, in the form presented by \cite{Gimenez1985} in degrees per orbital cycle as:
\begin{equation}
\label{eqrel2}
    \dot{\omega}_{\rm rel}=5.447\times10^{-4}\frac{\left(M_1+M_2\right)^{2/3}}{\left(1-e^2\right)P_{\rm a}^{2/3}} \; {\rm deg}\;{\rm cycle}^{-1},
\end{equation}
where $e$ is the orbital eccentricity, $M_1$ and $M_2$ are the component masses in solar units, and $P_{\rm a}$ is the anomalistic period in days, which measures the time between two consecutive periastron passages, and is related to the sidereal period, $P_{\rm s}$, through:
\begin{equation}
P_{\rm s}=P_{\rm a}\left(1-\frac{\dot{\omega}}{360}\right),
\end{equation}
where $\dot{\omega}$ is in degrees per orbital cycle. The classical contribution to the apsidal motion is produced by perturbations in the gravitational potential due to the lack of spherical symmetry in the shape of the components, which are distorted due to rotational flattening and tidal oblateness, the so-called quadrupole effect, and for the most part depends on the degree of mass concentration towards the centre \citep{Shakura1985}. The term $\dot{\omega}_{\rm cl}$, when only considering the contributions arising from the second-order harmonic distortions of the potential, can be described by the expressions given by \cite{Sterne1939} and \cite{Kopal1959}:
\begin{equation}
\label{eqclas1}
    \dot{\omega}_{\rm cl}=360\times \sum_{i=1}^2 \left( k_{2,i}c_i^{\rm rot}+k_{2,i}c_i^{\rm tid} \right)\; {\rm deg}\;{\rm cycle}^{-1},
\end{equation}
where the index $i$ refers to the component and $k_{2,i}$ are the second-order internal structure constants characterising the internal mass distribution of the stars, which can be derived from theoretical models by numerically integrating the Radau differential equation \citep{Kopal1978,Hejlesen1987,Schmitt2016}. The parameters $c_i^{\rm rot}$ and $c_i^{\rm tid}$ are the rotational and tidal contributions to the apsidal motion, respectively, which, assuming that the stellar and orbital rotation axes are aligned, can be expressed \citep{Kopal1978,Shakura1985} as:
\begin{equation}
\label{eqclas2}
   c_i^{\rm rot}= \frac{r^5_i}{(1-e^2)^2}\left(1+\frac{M_{3-i}}{M_i}\right)\left( \frac{\Omega_i}{\Omega_m} \right)^2,
\end{equation}
\begin{equation}
\label{eqclas3}
   c_i^{\rm tid}= 15r^5_i\frac{M_{3-i}}{M_i}\frac{\left(1+1.5e^2+0.125e^4\right)}{\left(1-e^2\right)^5},
\end{equation}
where $\Omega_m=2\pi/P_{\rm s}$ is the mean angular velocity of the orbital motion, $\Omega_i=v_{\rm rot,i}/R_i$ is the angular velocity of the rotation of each component, and $r_i=R_i/a$ are the relative component radii. In Eq. (\ref{eqclas1}) we do not use terms with $k_3$ and $k_4$ because they usually produce negligible contributions given the uncertainties of the lower order terms \citep{Claret1993,Rosu2020}. As all the terms appearing in Eq. (\ref{eqrel2}) can be obtained from observations, the GR apsidal motion rate is most often calculated analytically and subsequently subtracted from the measured rate to compare with stellar model predictions and provide constraints on interior structure \citep[e.g.][]{Claret1993,Khaliullin2007,Rauw2016,Zasche2019}.

Direct GR tests using measured apsidal motion rates have been hampered by the typical large relative uncertainty due to the error propagation from the dominant classical term. Eclipsing binaries suitable to test GR are those with relatively long orbital periods, where the general relativistic term makes up most of the apsidal motion contribution, that are still sufficiently short to produce an apsidal motion rate larger than the expected detection threshold  \citep[see equations 16 and 17 in][and their Table 1 to see the dependence of the period limits with the eccentricity]{Gimenez1985}. The most famous case is that of DI\,Her, whose detailed initial analyses showed disagreement with GR predictions \citep[e.g.][]{Guinan1985}. It was later shown \citep{Albrecht2009} that this mismatch was actually caused by a strong spin axis misalignment of the component stars, which leads to a different classical term, with a negative contribution of the rotational term to the total apsidal motion rate. There have been some attempts to further test GR using larger samples of eclipsing binaries but these failed to reach conclusive results \citep[e.g.][]{DeLaurentis2012}. This was caused by the large relative uncertainties resulting from the subtraction of the dominant classical contributions to the total apsidal motion term. The systems selected had a small fractional contribution of the GR term. With a carefully selected sample and the use of longer time baselines and more precise eclipse timings it should be possible to perform more accurate analyses and reach conclusive tests of GR. Precise determinations and even simply the detection of apsidal motion based on eclipse timings requires in most cases a dedicated long-term monitoring, generally spanning several decades. This is particularly difficult for systems whose orbital periods are long or have very slow apsidal motion rates, which happen to be the systems with the largest relativistic contributions due to the decrease of the classical terms with the orbital period as shown by \cite{Gimenez1985}.

\begin{table*}[t]
\centering
\caption{Properties of 15 eclipsing binary systems with eccentric orbits, accurate absolute dimensions, and literature-predicted relativistic apsidal motion relative contribution greater than 60\%.} \label{tab:systemsprop}
\begin{tabular}[t]{lllllllcl}
\hline
\hline
\noalign{\smallskip}
System    & $P_s$ [d]  & $e$ & $M_{1}$ [$M_{\odot}$]  & $M_{2}$ [$M_{\odot}$]     & $R_{1}$ [$R_{\odot}$]     & $R_{2}$ [$R_{\odot}$]  & $\% \dot{\omega}_{\rm Rel, Pred}$   & Ref. \\
\noalign{\smallskip}
\hline
\noalign{\smallskip}

KX\,Cnc   & 31.2197874(14)  & 0.4666(3) &       1.138(3)        & 1.131(3) &       1.064(2)        & 1.049(2) & 99 & Sow12\\

AL\,Dor   & 14.90537(2)  & 0.1952(2) &  1.1029(4)       & 1.1018(5) &   1.121(10)       & 1.118(10) & 98 &  G\&G\\

RW\,Lac   & 10.369205(2) & 0.0098(10) & 0.928(6)        & 0.870(4) &    1.186(4)        & 0.964(4) &  96 & Lac05 \\

V530\,Ori    & 6.1107784(3) & 0.0880(2) & 1.004(7) & 0.596(2) & 0.980(13) & 0.587(7) & 92 & Tor14 \\

HP\,Dra    & 10.7615(2) & 0.0367(9) &   1.133(5)        & 1.094(7) & 1.371(12) & 1.052(10) & 89 & Mil10\\

TZ\,Men     & 8.569000(10) & 0.035(3) & 2.49(3) & 1.504(10) & 2.02(2) & 1.432(15) & 86 & And87 \\

V541\,Cyg & 15.3378992(7) &     0.4684(14) & 2.335(17)  & 2.260(16) &   1.859(12)       & 1.808(15) & 86 &  Tor17\\

LV\,Her    & 18.435954(2) & 0.6127(7) & 1.193(10) & 1.170(8) &  1.358(12)       & 1.313(11) & 83 & Tor09 \\

V459\,Cas & 8.45825381(19) & 0.0244(4) & 2.02(3) & 1.96(3) & 2.009(13) & 1.965(13) & 80 & Lac04 \\

RR\,Lyn    & 9.94508(6) & 0.0793(9) & 1.927(8) & 1.507(4) & 2.57(2) & 1.59(3) & 80 & Tom06 \\

V501\,Her & 8.5976870(10)  & 0.0956(8) & 1.269(4) & 1.211(3) & 2.002(3) & 1.511(3) & 78 & Lac14\\

KW\,Hya & 7.750469(6) & 0.094(4) & 1.96(3) & 1.487(13) & 2.124(15) & 1.44(2) & 77 & G\&A \\

V501\,Mon & 7.0212077(10) & 0.1339(6) & 1.646(4) & 1.459(3) & 1.89(3) & 1.59(3) & 71 & Tor15 \\


GG\,Ori & 6.6314936(17) & 0.2218(22) & 2.342(16) & 2.338(17) & 1.85(3) & 1.83(3) & 67 & Tor00\\

EY\,Cep & 7.971488(6)  & 0.4429(14) & 1.523(8) & 1.498(14) & 1.463(10) & 1.468(10) & 64 & Lac06\\





\noalign{\smallskip}
\hline
\end{tabular}
\tablefoot{Only systems with primary and secondary eclipse coverage from the {\em TESS} mission are considered.}
\tablebib{Sow12: \cite{Sowell2012}; G\&G: \cite{Gallenne2019} and \cite{Graczyk2019}; Lac05: \cite{Lacy2005}; Tor14: \cite{Torres2014}; Mil10: \cite{Milone2010}; And87: \cite{Andersen1987}; Tor17: \cite{Torres2017}; Tor09: \cite{Torres2009}; Lac04: \cite{Lacy2004}; Tom06: \cite{Tomkin2006}; Lac14: \cite{Lacy2014}; G\&A: \cite{Gallenne2019} and \cite{Andersen1984}; Tor15: \cite{Torres2015}; Tor00: \cite{Torres2000}; Lac06: \cite{Lacy2006}.}
\end{table*}

Data from the Transiting Exoplanet Survey Satellite ({\em TESS}) mission \citep{Ricker2015} aimed at detecting exoplanets through photometric transits provide the opportunity to obtain densely-covered light curves of eclipsing binary systems at very high precision. Stellar eclipse events are generally much deeper than exoplanet transits and therefore {\em TESS} data are of excellent quality for such studies. Largely uninterrupted monitoring of the light curves is possible from space without the disturbing effect of the day-night cycle, therefore minimising the existence of phase gaps. In addition, most eclipsing binary systems were observed by {\em TESS} at the high cadence rate of 2 minutes. Thanks to all these circumstances, accurate eclipse timings can be obtained for numerous events. Furthermore, already during its initial 2 years of mission, {\em TESS} has covered a large fraction of the sky for a duration of at least 27 days, and therefore the chances of having observed eclipse events of most well-studied eclipsing binary systems are very high. Of course, the time baseline of the typical {\em TESS} observations is rather modest, but the high-precision measurements can be combined with past timings to enlarge the covered time-span significantly.

In this paper we perform apsidal motion determinations in eccentric eclipsing binaries with relatively long orbital periods and for which general properties such as mass, radius, and orbital period are accurately determined. We restrict our analysis to those systems with a literature-predicted relativistic contribution to the total apsidal motion rate of at least 60\%. In some cases, we are able to detect apsidal motion for the first time. Particular care is put in the estimation of the classical terms to determine the GR term of the apsidal motion rate to the best possible accuracy and with realistic uncertainties, meaning that a comparison with theory can be performed. We analyse the individual times of eclipse for 15 eclipsing binaries, from which we aim to determine a precise value of the GR apsidal motion term 
for each system. We then compare these terms with predictions from different gravitation theories. In Sect.\,\ref{sec:systems} we present the analysed sample and its general properties. In Sect.\,\ref{sec:minima} we explain the methodology used to determine the times of eclipse, while in Sect.\,\ref{sec:apsidal} we describe the methodology followed to determine the apsidal motion for each of the systems using {\em TESS} and archival timings. In Sect.\,\ref{sec:results} we present the resulting apsidal motion rates, the theoretical prediction of their classical contributions, and compare the differences with GR predictions. Finally, we use our measurements to test different gravitational theories in Sect.\,\ref{sec:test}, and provide our concluding remarks in Sect.\,\ref{sec:conclusions}.

\section{The eclipsing binary sample} \label{sec:systems}

For any useful interpretation of the observed apsidal motion in eccentric eclipsing binaries, it is essential to have good knowledge of the general properties of the component stars, mainly their masses and radii. Relevant equations such as Eqs.\,(\ref{eqclas2}) and (\ref{eqclas3}) have terms with high powers (up to the fifth) of the relative radii for example, thus enhancing their potential uncertainties. For this reason, we have limited our dynamical study to cases with well-studied and precise fundamental properties. The basic source is the compilation of well-studied detached eclipsing binaries by \cite{Torres2010}. This has been complemented with systems from the DEBCAT catalogue \citep{Southworth2015}, which is permanently updated, and includes eclipsing binaries with mass and radius measurements with 2\% accuracy or better, following the criterion in \cite{Andersen1991}.

From the compilations above and imposing the restriction of having {\em TESS} measurements for both primary and secondary eclipses, we selected the sample of eccentric eclipsing binaries in Table\,\ref{tab:systemsprop}. Given our objective of testing the theoretically predicted GR contribution, as given by Eq.\,(\ref{eqrel2}), with observed apsidal motion rates, we limited the sample to only systems with a literature-predicted fractional contribution greater than 60\%. We used the general properties, as referenced in Table\,\ref{tab:systemsprop}, and an estimated classical contribution given by Eq.\,(\ref{eqclas1}), with the assumption of rotational synchronisation at periastron and co-aligned rotational axes. Table\,\ref{tab:systemsprop} lists the adopted general properties of the systems such as the sidereal period, eccentricity, component masses, and radii of the 15 selected eclipsing binaries, together with the predicted GR contribution to the apsidal motion (\%$\dot{\omega}_{\rm Rel, Pred}$). It should be noted that in all tables presented throughout this paper the values in parentheses indicate the uncertainties affecting the last digits.

Three additional eclipsing binary systems not included in Table\,\ref{tab:systemsprop} deserve specific attention. DI\,Her is known to have misaligned rotational axes \citep{Albrecht2009}. The computation of the classical term of the apsidal motion therefore requires the use of the general form of Eq.\,(\ref{eqclas1}) given by \cite{Shakura1985} and an analysis of those angles constrained but not directly observed with the Rossiter-McLaughlin effect \citep{Claret2010}, which complicates the interpretation. 
EP\,Cru has component stars with rotational velocities 5.8 times higher than synchronisation at periastron \citep{Albrecht2013}. When such rotational velocities are considered, the classical term becomes dominant and the system does not meet our limiting criterion on the fractional GR contribution to the total apsidal motion rate to be included in the present study. Finally, a detailed analysis of BF\,Dra revealed a trend in the primary and secondary eclipse residuals of the high-precision {\em TESS} timings. Such a trend may likely be produced by the presence of a third body orbiting the system. The unconstrained nature of the third companion makes the determined apsidal motion rate ill-suited for comparison with theoretical predictions. Eclipsing binaries with a relative contribution of the relativistic term of below 60\%, together with DI\,Her, EP\,Cru, and BF\,Dra, will be analysed in a subsequent paper focusing on the classical terms (tidal and rotational).

\section{Determination of times of minimum light} \label{sec:minima}

We used {\em TESS} data from sectors 1 to 29 to compute precise timings of the primary and secondary eclipses for all of our targets. We gathered the 2-min short-cadence simple aperture photometry (SAP) produced by the Science Process Operation Centre \citep[SPOC,][]{Jenkins2016} available at the Mikulski Archive for Space Telescopes\footnote{https://mast.stsci.edu/portal/Mashup/Clients/Mast/Portal.html}. In the cases of V530\,Ori and V501\,Mon, for which the 2-min cadence photometry was not available, we extracted the 30-min cadence simple aperture photometry from the {\em TESS} full frame images (FFIs) using the public {\em TESS} aperture photometry tool \textsc{Eleanor}\footnote{https://adina.feinste.in/eleanor/} \citep{Feinstein2019}. Table\,\ref{tab:TESS} lists the {\em TESS} sectors and cadence at which each system has been observed, the time-span between the first and last eclipse considered for each target, and their {\em Gaia} $G$ magnitudes \citep{Gaia2016,Gaia2018}. Possible systematic deviations present in {\em TESS} data may include activity-induced modulations or other geometric effects that could bias the measurement of the time of minimum. To mitigate these effects, we employed the python package george \citep{Foreman2015} to model the out-of-eclipse photometry using a Gaussian process correlated-noise model with a squared-exponential covariance function \citep[see e.g.][for more details]{Gibson2011,Aigrain2016}. The resulting model was used to normalise the entire light curve, including the eclipses. The length-scale hyperparameter was constrained to values above twice the duration of the eclipse to avoid adding spurious high-frequency noise inside the eclipse region. 

\begin{table}[t]
\centering
\caption{{\em TESS} sectors and cadence at which the sample of studied eclipsing binary systems have been observed, the time-span between the first and last eclipse included in this work, and their {\em Gaia} $G$ magnitudes.} \label{tab:TESS}
\begin{tabular}[t]{lcrrc}
\hline
\hline
\noalign{\smallskip}
\multirow{2}{*}{System}   & \multirow{2}{*}{Sector} & Cad. & $\Delta t$ & $G$\\
  & & [min] & [d] & [mag]\\
\noalign{\smallskip}
\hline
\noalign{\smallskip}

KX\,Cnc   & 21 & 2 & 20 & 7.0363(2)  \\

AL\,Dor  & 1--6, 8--13, 27--29 & 2 & 783 & 7.5820(3) \\

RW\,Lac  & 16, 17 & 2 & 47 &  10.4381(3)\\

V530\,Ori & 6 & 30 & 18 & 9.6959(4) \\

HP\,Dra  & 14, 15, 26 & 2 & 344 & 7.8117(4) \\

TZ\,Men  & 12, 13, 27 & 2 & 428  & 6.1556(8) \\

V541\,Cyg  & 14 & 2 & 22 & 10.3286(4)\\

LV\,Her  & 25, 26 & 2 & 34 & 10.8853(6) \\


V459\,Cas   & 18, 24 & 2 & 186  & 10.2905(3)\\

RR\,Lyn   & 20 & 2 & 20 & 5.4696(14) \\

V501\,Her & 25, 26 & 2 & 47  & 10.9787(6) \\

KW\,Hya  & 8 & 2 & 20 &  6.0529(7)\\

V501\,Mon   & 6 & 30 & 17  &  12.1884(2)\\



GG\,Ori  & 6 & 2 & 16 & 10.2372(5)\\

EY\,Cep  & 19, 25 & 2 & 186 & 9.7154(3)\\





\noalign{\smallskip}
\hline
\end{tabular}
\end{table}

\begin{figure}[t]
\centering
\includegraphics[width=\columnwidth]{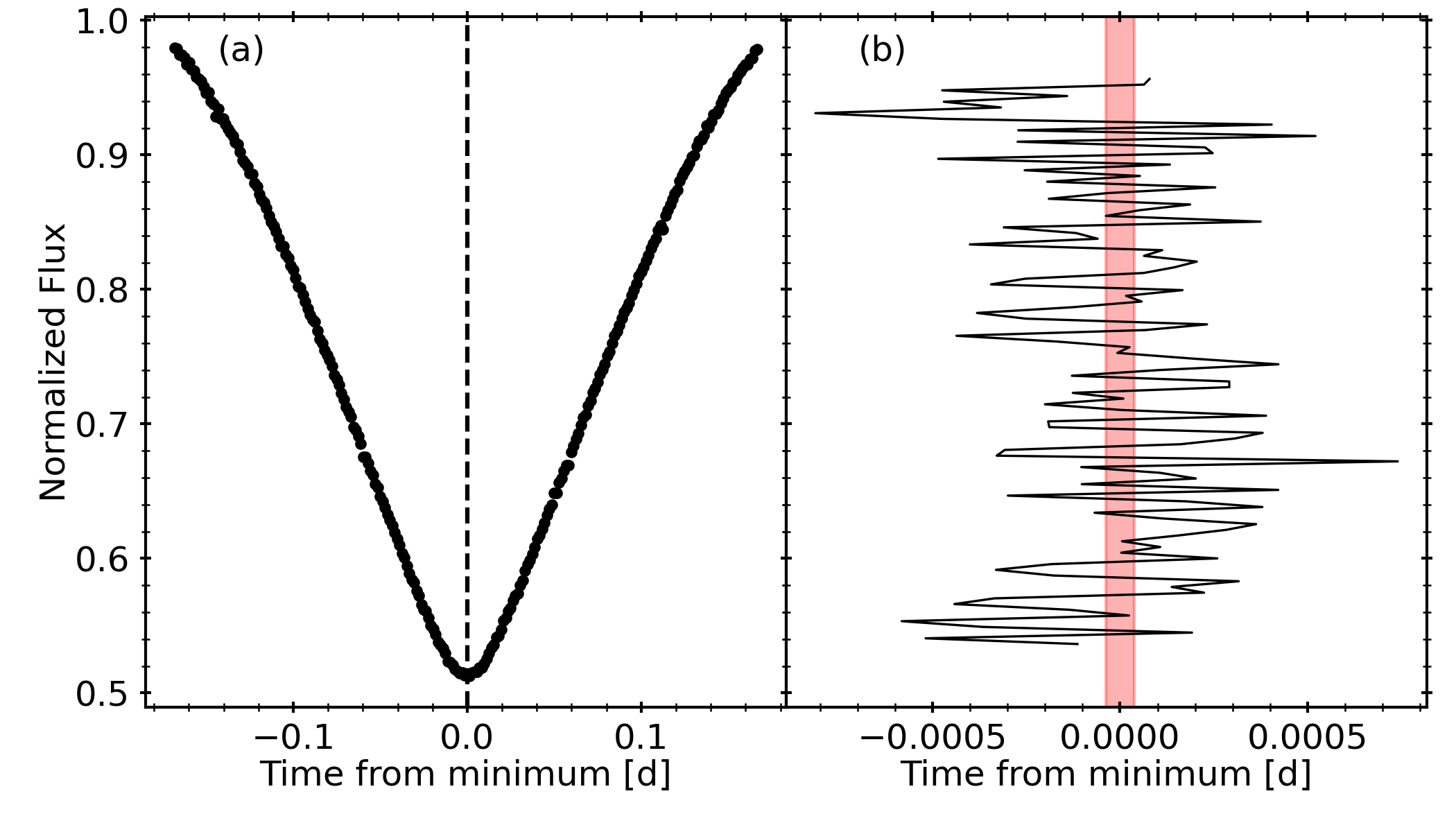}
\includegraphics[width=\columnwidth]{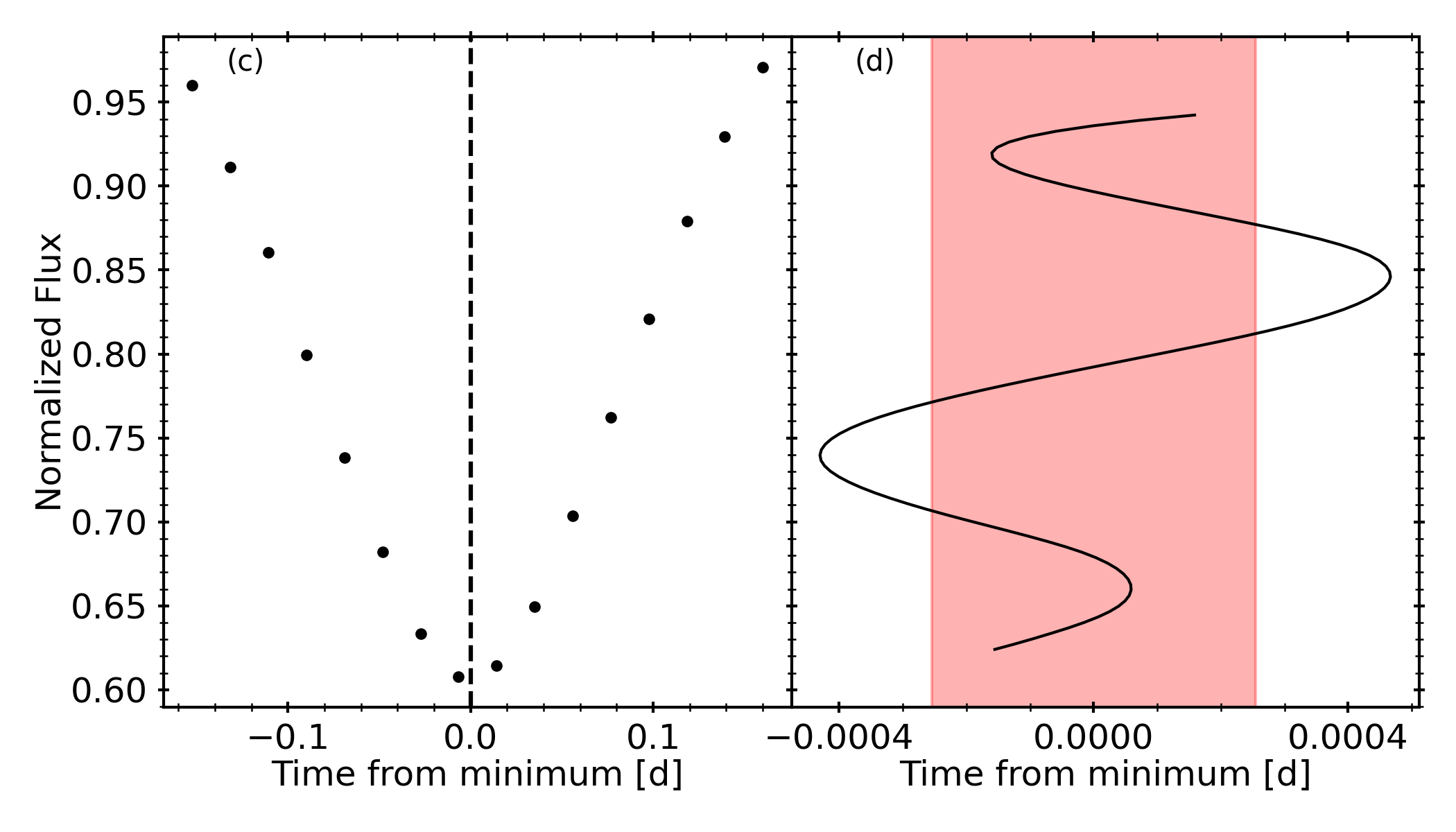}
\caption{Examples of {\em TESS} data for an eclipse of V459\,Cas \textit{(a)} with 2-min cadence, and V501\,Mon \textit{(c)} with 30-min cadence. In panels \textit{(b)} and \textit{(d)} the black line shows the bisector corresponding to each eclipse.  The computed time of minimum for each system, which is used as reference time in the plots, is marked as a black-dashed line in panels (a) and (c). Its uncertainty is illustrated in panels (b) and (d) as a read shaded area. The resulting timing uncertainty is 3.3\,sec for V459\,Cas and 22.0\,sec for V501\,Mon.} 
          \label{fig:bis}%
\end{figure}

The \cite{Kwee1956} method (hereafter, KvW) was adopted to determine all the times of minimum light. For consistency, we used the same orbital phase interval for all primary and secondary eclipses, and we only considered eclipses for which ingress and egress is well sampled. The median uncertainty in the obtained times of minimum are 2.5 and 12.9 seconds for the 2-min and 30-min cadence photometry, respectively. The precision of the KvW method for an equally sampled ideal eclipse is inversely proportional to the square root of the number of points used. Given that typically the number of {\em TESS} photometric points in the 2-min cadence eclipses is 15 times larger than in the 30-min cadence eclipses, we expect an error ratio between the two cadences of about $\sqrt{15}$, as approximately observed. In Fig.\,\ref{fig:bis} we show two examples of eclipses observed with {\em TESS} using 2-min and 30-min cadence observations and we mark the time of minimum computed using the KvW method.

We computed the bisector of each eclipse to diagnose local asymmetries that could be caused by stellar activity for example. Bisector points were determined as the time average between two symmetric points of the ingress and egress branches. Cubic spline interpolation was used to define points of equal flux. From the calculated values, we defined two indicators. The $bis_{odd}$ parameter is defined as the difference of the average bisector between 50--100\% and 0--50\% of the eclipse depth (i.e. $<bis_{50-100}> - <bis_{0-50}>$), while the $bis_{even}$ parameter is computed as the difference between the average of the two intervals at the top 75--100\% and the bottom 0--25\%, and the average of the middle 25--75\% of the eclipse depth (i.e. $<bis_{75-100},bis_{0-25}> - <bis_{25-75}>$). These bisector parameters represent first- and second-order measures of distortions in the eclipse shapes. We used them as signposts for the potential presence of biased timing determinations,  for example those due to issues with the data or effects on the stellar surface such as spots. As a result, we did not consider for further analysis those times of minimum with one of these indices deviating by more than 3\,$\sigma$ from their mean. Table\,\ref{tab:min} provides all the measured times of minimum light, the type of eclipse, and the bisector indicators $bis_{odd}$ and $bis_{even}$. Also, the right panel of Fig.\,\ref{fig:bis}  shows two examples of the bisectors of two eclipses and the uncertainty in the determination of their time of minimum.

In the eclipses analysed in this work the final measurement errors are 1 to 11 times lower than the dispersion of the individual points used to measure the bisector. Given that the number of photometric points comprised within the eclipses ranges from 8 to 440, that the bisector points are computed from two photometric points, and that the timing error for an ideal eclipse scales with $N^{-1/2}$, we expect errors that are 2 to 15 times lower than the bisector dispersion (i.e. $\sqrt{8/2}$ and $\sqrt{440/2}$). When applied to our data, this methodology yields uncertainties that are larger than those expected for an ideal eclipse, which may be taken as an indication that the errors are not underestimated. Potential systematic deviations from an ideal eclipse that are not accounted for by the KvW method caused several works to consider the error estimates coming from this method as exceedingly optimistic \citep{Breinhorst1973,Mikulasek2014}. However, with the novel use of the bisectors as indicators for such deviations, we were able to detect the affected eclipses and discard them for further use, therefore ensuring that our measurement errors are a good representation of the true statistical precision.


\begin{table}[t]
\centering
\caption{Computed {\em TESS} times of minimum light and bisector indicators of the systems in Table\,\ref{tab:systemsprop}.} \label{tab:min}
\begin{tabular}[t]{llcrr}
\hline
\hline
\noalign{\smallskip}
System    & $T_0$ [BJD]  & Type  & $bis_{odd}$ [s] & $bis_{even}$ [s]  \\\noalign{\smallskip}
\hline
\noalign{\smallskip}
KX\,Cnc &   2458876.936865(7) & 1 & $-$0.1 & $-$1.4 \\
        &  2458897.01419(4) & 2 & 29.4  & 24.3 \\
AL\,Dor &   2458330.899178(14) & 2 & $-$2.5 &    2.0 \\
        &  2458345.804510(17) & 2 & 1.7   & 1.6 \\
        &  2458360.709872(15) & 2 & $-$0.2 & $-$0.9 \\
\hline
\end{tabular}
\tablefoot{This table is presented in its entirety in the online version of the article.}
\end{table}

\section{Apsidal motion determinations} \label{sec:apsidal}

Several methods can be employed to estimate apsidal motion rates in eclipsing binaries, such as the changing position of the eclipses with respect to each other over time \citep[e.g.][]{Gimenez1987,Wolf2006,Wolf2010,Kim2018}, the change in the shape of the radial velocity curve over time \citep[e.g.][]{Ferrero2013,Schmitt2016,Rauw2016}, or the change in the shape of the light curve over time \citep{Bakis2008,Harmanec2014,Torres2017}. Here we use the equations of \cite{Gimenez1995} defining the times of eclipse:
\begin{equation}
\begin{split}
  T_j= & T_0+NP_{\rm a}\left(1-\frac{\dot{\omega}}{360}\right)+(j-1)\frac{P_{\rm a}}{2}+\\ & +\sum_{i=1}^{3}(-1)^{i-1}A_{2i}\frac{P_{\rm a}e^{2i}}{2^{2i}\pi}\sin(2i\omega) \: + \\
     & +\sum_{i=0}^{2}(2j-3)A_{2i+1}\frac{P_{\rm a}e^{2i+1}}{2^{2i+1}\pi}\cos[(2i+1)\omega],
 \end{split}
 \label{eqt0}   
\end{equation}
where $\dot{\omega}$ is expressed in deg\,cycle$^{-1}$, $j$ is 1 or 2 for the primary or secondary eclipse, respectively, $T_0$ is the time of eclipse at epoch $N=0$, $P_{\rm a}$ is the anomalistic period, and $\omega$ is the argument of periastron expressed
in radians at time $T$, which depends on $\dot{\omega}$ and the argument of periastron at time $T_0$, $\omega_0$, as $\omega=\omega_0+N\dot{\omega}$. Finally, the coefficients $A_i$ depend on the eccentricity $e$ and the inclination of the orbit, and their full expressions are given in eqs. (16) to (21) of \cite{Gimenez1995}.

Equation\,(\ref{eqt0}) is complete up to $\mathcal{O}(e^5)$ \citep{Gimenez1995}, and is therefore only valid for orbital eccentricities that do not reach extremely high values, that is, below 0.7. Fitting this equation to the actual individual measurements allows the determination of the apsidal motion rate, $\dot{\omega}$, as well as the orbital eccentricity, $e$, the orbital period, the reference time of eclipse, $T_0$, and the argument of periastron at that time, $\omega_0$. It should be noted that there are significant differences in the calculated $\dot{\omega}$ when only expressions considering low orders in $e$ are used. The severity of the deviations is a function of the eccentricity but we use terms up to $e^5$ for all cases, implying relative errors well below 1\%.

Strong degeneracies between $e$ and $\omega_0$ can limit the use of Eq.\,(\ref{eqt0}) to derive all parameters simultaneously unless a significant fraction of the apsidal motion period is covered with observational data. Attempts at five-parameter simultaneous fits using a poorly covered apsidal period may lead to biased solutions because of resulting correlations between $e$ and $\omega$. Potential issues are further exacerbated by the possible presence of light-time effect variations caused by an orbiting third body. To obtain non-degenerate results when the time-span of the observations is limited, the value of the eccentricity, which is directly related to the amplitude of the variations, should at least be adopted independently, for example from modelling the light and radial velocity curves. When the eccentricity is fixed, the instantaneous shape of the apsidal motion curve can be described by the tangent at $\omega_0$, which can also be expressed in terms of the difference between the periods corresponding to primary and secondary eclipses, as described by \cite{Gimenez1995}, considering that the value of $\omega$ at the time of the eclipses does not differ from $\omega_0$. This is just the time-derivative of the difference between the times of secondary ($T_2$) and primary ($T_1$) eclipses:
\begin{equation}
 \frac{d(T_2-T_1)}{dt}=\frac{\dot{\omega}P_{\rm a}}{180}\left( \sum_{i=0}^{2}(-1)^{i}A_{2i+1}(2i+1)\frac{e^{2i+1}}{2^{2i}}\sin[(2i+1)\omega_0] \right),
    \label{eqt2t1}
\end{equation}
with $\dot{\omega}$ expressed in deg\,cycle$^{-1}$.

In our study, we use the values of $T_2-T_1$ from pairs of close primary and secondary minima of high-precision data. This is done to ensure maximum robustness of the apsidal motion determinations and to minimise the possible effects over the measured orbital period that might be produced by systematic errors arising from different methodologies or astrophysical variability. We favour this method over the use of the difference between periods resulting from linear fits to individual primary and secondary eclipses. This is equivalent to the right term of Eq. (\ref{eqt2t1}). Using $T_2-T_1$ from close eclipses has several advantages, namely that these are more likely to be obtained by the same observer, that both timings correspond to the same epoch of stellar activity, and that it avoids effects from light-travel time caused by a potential third body. However, it should be noted that the presence of a third body may induce other perturbations such as the eccentric Lidov-Kozai effect \citep{Lidov1962,Kozai1962,Naoz2016} that could alter the observed apsidal motion rates. A full characterisation of the system is needed to take into account the effects of a potential third body. In the absence of such a detailed analysis, a comparison with the theoretically predicted apsidal motion rate could be affected. 

The downside our adopted approach is that the number of $T_2-T_1$ measurements ({\em TESS} plus archival) for some of the studied systems presented below is rather low. Depending on the time baseline and the impact of effects such as stellar activity, this may lead to some unaccounted-for source of additional error. However, our entire procedure has been devised to minimise such effects (strict selection of minimum timings, bisector correction criteria, etc.) and therefore we do not expect systematic deviations to significantly affect our measurements. Also, most of the eclipsing binaries that we analyse here do not show signs of any stellar activity, either because they lack a convective envelope or because of their low rotational velocities.

The eclipsing binaries in our sample have been typically monitored between {\em TESS} and archival data for less than 1\% of their apsidal motion period, of the order of thousands of years. Therefore, the observed $T_2-T_1$ values should accurately define the tangent to the curve described by Eq.\,(\ref{eqt2t1}), and relying on a previously determined orbital eccentricity, Eq.\,(\ref{eqt2t1}) is fully applicable. The approximation of the derivative is of course sensitive to the argument of periastron and its precision decreases for values of $\omega_0$ near 0 or 180 degrees, when the variation of $T_2-T_1$ over time and the corresponding period differences become close to zero. We assume the eccentricity derived from the best light and radial velocity curves, with the corresponding argument of periastron, in order to obtain the apsidal motion rate. We had to restrict the analysis to systems with high-precision data, as already discussed, and we focus our analysis on the systems listed in Table\,\ref{tab:systemsprop}, which are discussed individually below. 

All obtained secondary minus primary eclipse timings, $T_2-T_1$, for each of the available primary eclipses are given in Table\,\ref{tab:a1} with the corresponding uncertainty computed from the propagation of errors of individual timings, given in parentheses. We also list the orbital cycle ($N$) computed with respect to the reference time $T_0$ given in the first column. For those primary eclipses without any secondary counterpart on the same orbital cycle, we computed the secondary minus primary eclipse timings as $T_2-T_1-dN\times P_{\rm s}$, where $P_{\rm s}$ is the sidereal period listed in Table\,\ref{tab:systemsprop} and $dN$ is the difference in orbital cycles between the two minima. We list $dN$ in the last column in Table\,\ref{tab:a1}. 

Below we discuss the studied systems in detail and describe the methodology applied to determine their apsidal motion rates in those nine cases where such a measurement was possible. For 6 out of the 15 systems in Table\,\ref{tab:systemsprop}, namely, RW\,Lac, V530\,Ori, HP\,Dra, TZ\,Men, LV\,Her, and RR\,Lyn, we could not find a measurable apsidal motion rate in spite of having high-precision {\em TESS} data available, but we have nevertheless included their time of minima and $T_2-T_1$ values in Tables \ref{tab:min} and \ref{tab:a1}, which could be valuable for future determinations. For the systems in which we could determine an apsidal motion rate, additional values needed for their analysis, such as the relative radii or the projected rotational velocities, are listed in Table\,\ref{tab:otherprop}, together with the values of $\omega_0$ and the reference time $T_0$.

\begin{table}[p]
\centering
\caption{$T_2-T_1$ computed from {\em TESS} light curves. The BJD value below each system name defines the origin epoch of the orbital cycle count ($N$).}
\label{tab:a1}
\begin{tabular}{lrcr} 
\hline\hline
\noalign{\smallskip}
System \& $T_0$ & $N$ & $T_2-T_1$ [d]  & $dN$ \\
\noalign{\smallskip}
\hline
\noalign{\smallskip}
KX\,Cnc  & \multirow{2}{*}{63} & \multirow{2}{*}{20.077329(36)} & \multirow{2}{*}{0}\\
$2456910.0768$  & & & \\
\hline
\noalign{\smallskip}
AL\,Dor   & 0 & 6.887539(24) & 0\\
$2458368.7277$ & 2 & 6.887447(18) & 0\\
         & 3 & 6.887438(22) & $-$1\\
         & 4 & 6.887435(17) & 0\\
         & 5 & 6.887471(19) & $-$1\\
         & 6 & 6.887408(20) & 1\\
         & 7 & 6.887431(18) & 0\\
         & 8 & 6.887474(20) & $-$1\\
         & 10 & 6.887478(19) & 0\\
         & 12 & 6.887378(18) & 0\\
         & 13 & 6.887413(22) & $-$1\\
         & 14 & 6.887361(23) & 1\\
         & 15 & 6.887402(24) & 0\\
         & 16 & 6.887392(25) & $-$1\\
         & 17 & 6.887326(22) & 1\\
         & 18 & 6.887382(22) & 0\\
         & 19 & 6.887416(23) & 0\\
         & 20 & 6.887391(16) & 0\\
         & 21 & 6.887451(20) & $-$1\\
         & 45 & 6.887292(21) & 0\\
         & 46 & 6.887294(22) & $-$1\\
         & 47 & 6.887226(17) & 0\\
         & 48 & 6.887218(19) & 0\\
         & 49 & 6.887235(21) & 0\\
         & 50 & 6.887228(24) & $-$1\\
\hline
\noalign{\smallskip}
RW\,Lac   & 0 & 5.111455(90) & $-$1\\
$2458744.8166$ & 1 & 5.111474(75) & 0\\
         & 2 & 5.11141(10) & 0\\
         & 4 & 5.111422(82) & $-$1\\
\hline
\noalign{\smallskip}
V530\,Ori & 0 & 2.83585(35) & 0\\
$2458471.0870$ & 2 & 2.83657(38) & 0\\
         & 3 & 2.83681(53) & $-$1\\
\hline
\noalign{\smallskip}
HP\,Dra  & 0 & 5.568563(60) & 0\\
$2458692.9375$ & 1 & 5.568771(48) & 0\\
         & 2 & 5.568480(55) & 0\\
         & 4 & 5.56850(19) & $-$1\\
         & 30 & 5.568262(58) & 0\\
         & 31 & 5.568604(36) & 0\\
\hline
\noalign{\smallskip}
TZ\,Men  & 0 & 4.371606(67) & 0\\
$2458633.4292$   & 1 & 4.371309(52) & 0\\
         & 2 & 4.371558(31) & 0\\
         & 3 & 4.371976(35) & 0\\
         & 5 & 4.371702(25) & 0\\
         & 48 & 4.371928(53) & $-$1\\
         & 49 & 4.371867(18) & 0\\
\hline
\noalign{\smallskip}
V541\,Cyg & 513 & 7.032672(81) & 1\\
$24550817.9760$ & 514 & 7.032794(84) & 0\\
\hline
\noalign{\smallskip}
LV\,Her & 0 & 15.898401(63) & 0\\
$2458998.6220$  & 1 & 15.898531(70) & 0\\
\hline
\noalign{\smallskip}
V459\,Cas  & 471 & 4.164664(60) & 0\\
$2454815.5698$  & 472 & 4.164493(53) & 0\\
         & 490 & 4.164565(70) & 0\\
         & 492 & 4.164777(56) & 0\\
\hline
\noalign{\smallskip}
RR\,Lyn  & 0 & 4.478175(71) & $-$1\\
$2458851.9265$   & 1 & 4.478547(46) & 0\\
\hline
\noalign{\smallskip}
V501\,Her & 388 & 4.120736(96) & 0\\
$2455648.5943$         & 389 & 4.121067(84) & 0\\
\hline
\end{tabular}
\end{table}

\addtocounter{table}{-1}
\begin{table}[t]
\centering
\caption{Continued.}
\begin{tabular}{lrcr} 
\hline\hline
\noalign{\smallskip}
System \& $T_0$ & $N$ & $T_2-T_1$ [d]  & $dN$ \\
\noalign{\smallskip}
\hline
\noalign{\smallskip}
         & 390 & 4.121030(76) & 0\\
         & 392 & 4.12087(10) & $-$1\\
         & 393 & 4.120252(98) & 0\\
\hline
\noalign{\smallskip}
KW\,Hya  & 657 & 3.549856(13) & 0\\
$2453430.9533$  & 658 & 3.549921(16) & $-$1\\
         & 659 & 3.549835(23) & $-$2\\
\hline
\noalign{\smallskip}
V501\,Mon   & -1 & 3.14676(16) & 0\\
$2458477.9808$ & 1 & 3.14675(29) & 0\\
\hline
\noalign{\smallskip}
GG\,Ori  & 983 & 2.790712(33) & 0\\
$2451952.4777$ & 985 & 2.790716(35) & 0\\
\hline
\noalign{\smallskip}
EY\,Cep  & -2 & 3.140134(29) & 0\\
$2458836.4248$ & 0 & 3.140225(32) & 0\\
         & 19 & 3.139902(31) & 0\\
         & 21 & 3.139882(35) & 0\\
\hline
\end{tabular}
\end{table}

\begin{table*}[t]
\centering
\caption{Other properties used to compute the apsidal motion of the systems analysed, such as the reference time and the corresponding argument of periastron, the stellar relative radii and projected rotational velocities, and the orbital inclination of the system.} 
\label{tab:otherprop}
\begin{tabular}{llrllrrl} 
\hline\hline
\noalign{\smallskip}
\multirow{2}{*}{System} &  \multicolumn{1}{c}{$T_0$} &  \multicolumn{1}{c}{$\omega_0$} &  \multicolumn{1}{c}{\multirow{2}{*}{$r_1$}} &  \multicolumn{1}{c}{\multirow{2}{*}{$r_2$}} &   \multicolumn{1}{c}{$v_1\sin i$} &  \multicolumn{1}{c}{$v_2\sin i$} &  \multicolumn{1}{c}{$i$} \\
 &  \multicolumn{1}{c}{[BJD]} &  \multicolumn{1}{c}{[deg]} &  &  &   \multicolumn{1}{c}{[km\,s$^{-1}$]} &  \multicolumn{1}{c}{[km\,s$^{-1}$]} &  \multicolumn{1}{c}{[deg]}\\
\noalign{\smallskip}
\hline
\noalign{\smallskip}
KX\,Cnc & 2456910.0768 & 63.76 & 0.01940(4) & 0.01913(4) &  6.4(1.0)  & 6.5(1.0) & 89.825(3) \\
AL\,Dor & 2458368.7277 & 107.45 & 0.0333(3) & 0.0334(3) &  4.6(1.0)  & 4.6(1.0) & 88.79(11)\\
V541\,Cyg & 2450817.9760 & 262.72 & 0.0431(2) & 0.0419(3) &  15(1) & 15(1) & 89.83(3)\\
V459\,Cas & 2454815.5698 & 240.32 & 0.0726(3) & 0.0710(3) &  54(2) & 43(2) & 89.467(7)\\
V501\,Her & 2455648.5943 & 250.10 & 0.08375(11) & 0.06323(11) &  14.3(1.0)      & 12.5(1.0) & 89.11(4)\\
KW\,Hya & 2453430.9533 & 225.29 & 0.0853(5) & 0.0594(8) & 15(2) & 13(2) & 87.5(3)\\
V501\,Mon & 2458477.9808 & 232.79 & 0.0838(13) & 0.0707(13) &  16.5(1.0) & 12.4(1.0) & 88.02(8)\\
GG\,Ori & 2451952.4777 & 122.89 & 0.0746(10) & 0.0737(10) &  24(2) & 23(2) & 89.30(10)\\
EY\,Cep & 2458836.4248 & 110.14 & 0.0603(4) & 0.0606(8) &  10(1) & 10(1) & 89.89(3)\\

\noalign{\smallskip}
\hline
\end{tabular}
\end{table*}

\subsection{KX\,Cnc}

The general properties of this highly eccentric system ($e=0.4666\pm0.0003$), such as mass, radius, eccentricity, and period, shown in Table \ref{tab:systemsprop}, are the result of the combined analysis of the light and radial velocity curves carried out by \cite{Sowell2012}, who reported no apsidal motion. The {\em TESS} data listed in Table\,\ref{tab:a1} yield $T_2-T_1 = 20.077329 \pm 0.000036$\,days. This measurement can be compared with $T_2-T_1$ values from the literature to search for the presence of apsidal motion. Some published minima by \cite{Davies2007} have insufficient precision and we therefore considered the light curve obtained by \cite{Sowell2012}. The authors give a phase for the secondary eclipse of 0.64325, which is not measured directly but results from the joint analysis of the light curve and the radial velocity curve. We therefore decided to retrieve the original photometric data, \textit{b}- and \textit{y}-band in the Stromgren system, and calculated the position of the eclipses using the same method as for the {\em TESS} measurements, yielding a value of $T_2-T_1 = 20.08076 \pm 0.00015$\,days, which is equivalent to a phase of the secondary of 0.64321. Comparing the value of $T_2-T_1$  with the {\em TESS} results, we observe a decrease of $-0.00343 \pm 0.00015$\,days after 177 orbital cycles, indicating a measurable apsidal motion. With only two data points available, an analytical derivation of the statistical uncertainty in the slope could be imprecise and biased. Therefore, we performed $10^6$ Monte Carlo simulations of $T_2-T_1$, following a Gaussian distribution with the corresponding uncertainty, and computed the standard deviation of all the solutions. From the resulting slope, and combining it with the eccentricity given by \cite{Sowell2012} from their joint solution, an apsidal motion rate of $\dot{\omega} = 0.000131 \pm 0.000010$\,deg\,cycle$^{-1}$ is derived, including the uncertainty in the adopted orbital parameters.

\subsection{AL\,Dor}

\begin{figure}[t]
\centering
\includegraphics[width=\columnwidth]{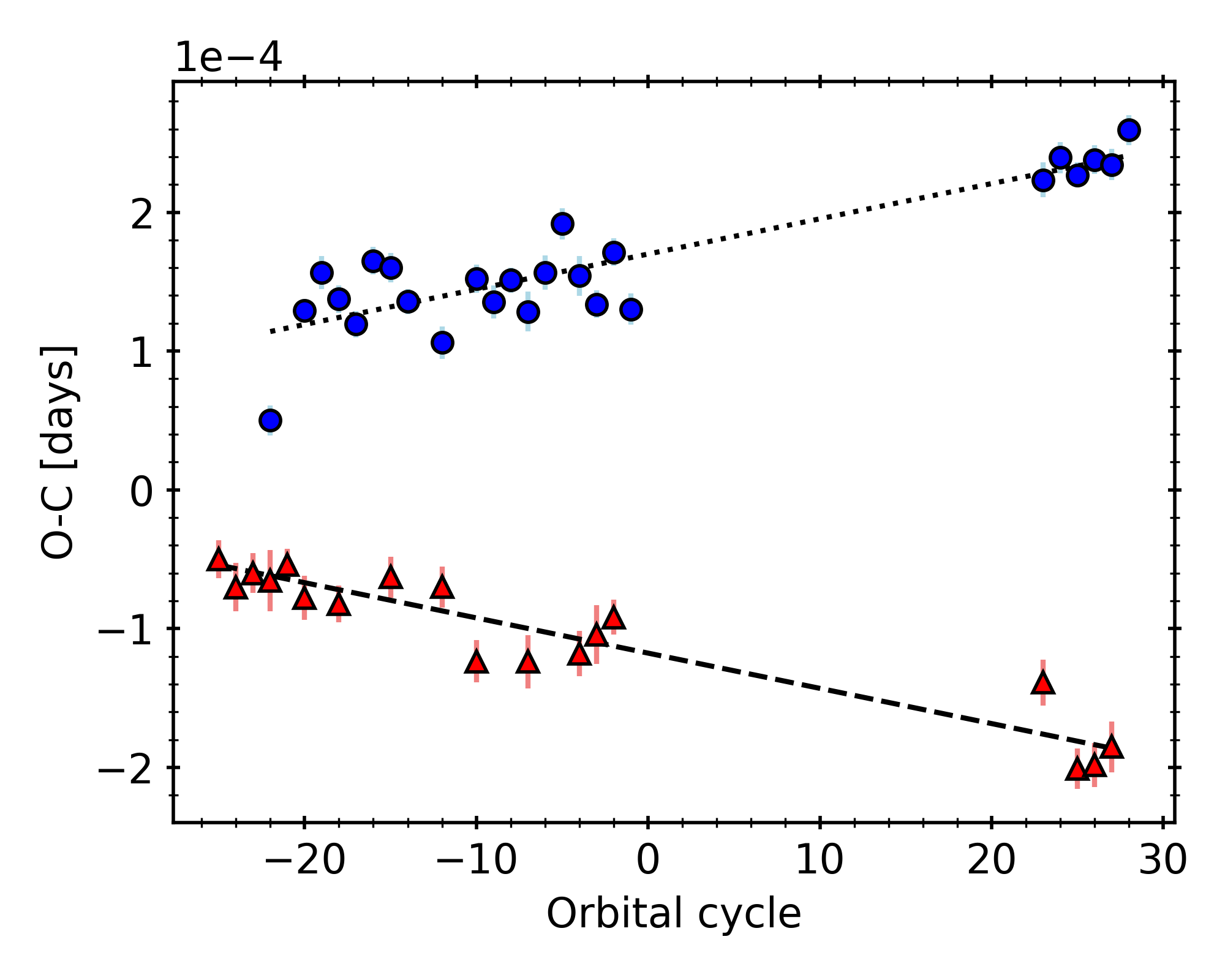}
\caption{Ephemeris curve for AL\,Dor as a function of the orbital cycle. The dotted and dashed lines represent fits to the primary (blue circles) and secondary (red triangles) eclipses, respectively. For a better visualisation, an arbitrary shift to the primary and secondary minima has been applied.}
          \label{fig:ALDor}%
\end{figure}

The general properties given in Table\,\ref{tab:systemsprop} are derived from the radial velocity curve of \cite{Graczyk2019}, combined with the astrometric solution by \cite{Gallenne2019}. No indication of apsidal motion was reported. The adopted parameters, including the sidereal period, are not based on a light-curve solution. Fortunately, our {\em TESS} eclipse timings, shown in Table\,\ref{tab:min}, yield a good number of accurate minima values indicating a slow apsidal motion without any need for archival measurements, as illustrated in Fig.\,\ref{fig:ALDor}. We tried to enlarge the time baseline to improve the apsidal motion determination, but we could not find earlier data of sufficient precision. The linear weighted fit to the $T_2-T_1$ values listed in Table\,\ref{tab:a1} corresponds to a slope of $-(5.0 \pm 0.5) \times 10^{-6}$\,days\,cycle$^{-1}$. The uncertainty of the sidereal period is large compared to the errors of periods based on a light-curve solution. For this reason we only used values of $T_2-T_1$ within the same orbital cycle to eliminate the effect of the uncertainty in the sidereal period. Complementarily, we considered all the individual timings to fit the periods corresponding to primary and secondary eclipses, obtaining a difference between them of $-(5.08 \pm 0.19) \times 10^{-6}$\,days\,cycle$^{-1}$, in agreement with our value based on measurements within the same orbital cycle. The eccentricity given by \cite{Gallenne2019}, $e = 0.1952 \pm 0.0002$, produces a final value of $\dot{\omega} = 0.000163 \pm 0.000006$\,deg\,cycle$^{-1}$.

\subsection{V541\,Cyg}

Apsidal motion in V541\,Cyg was measured by \cite{Khaliullin1985} and revised by \cite{Volkov1999}. \cite{Wolf2010} used the individual times of minimum light to obtain $\dot{\omega}$ = $0.00032\pm0.00006$\,deg\,cycle$^{-1}$, leaving the eccentricity as a free parameter, while \cite{Kim2018} obtained $\dot{\omega} = 0.000397 \pm 0.000013$\,deg\,cycle$^{-1}$, fixing the orbital eccentricity to the value of 0.479 given by \cite{Lacy1998}. \cite{Torres2017} obtained a new radial velocity curve and reanalysed the $V$-band light curve of \cite{Khaliullin1985}. For the apsidal motion determination, \cite{Torres2017} carried out a global analysis of all the data, photometric and spectroscopic, with variable argument of periastron. In this way, the authors determined the optimal value of the eccentricity to be $e = 0.4684 \pm 0.0014$ and an apsidal motion rate of $\dot{\omega} = 0.000360 \pm 0.000012$\,deg\,cycle$^{-1}$. 

\begin{figure}[t]
\centering
\includegraphics[width=\columnwidth]{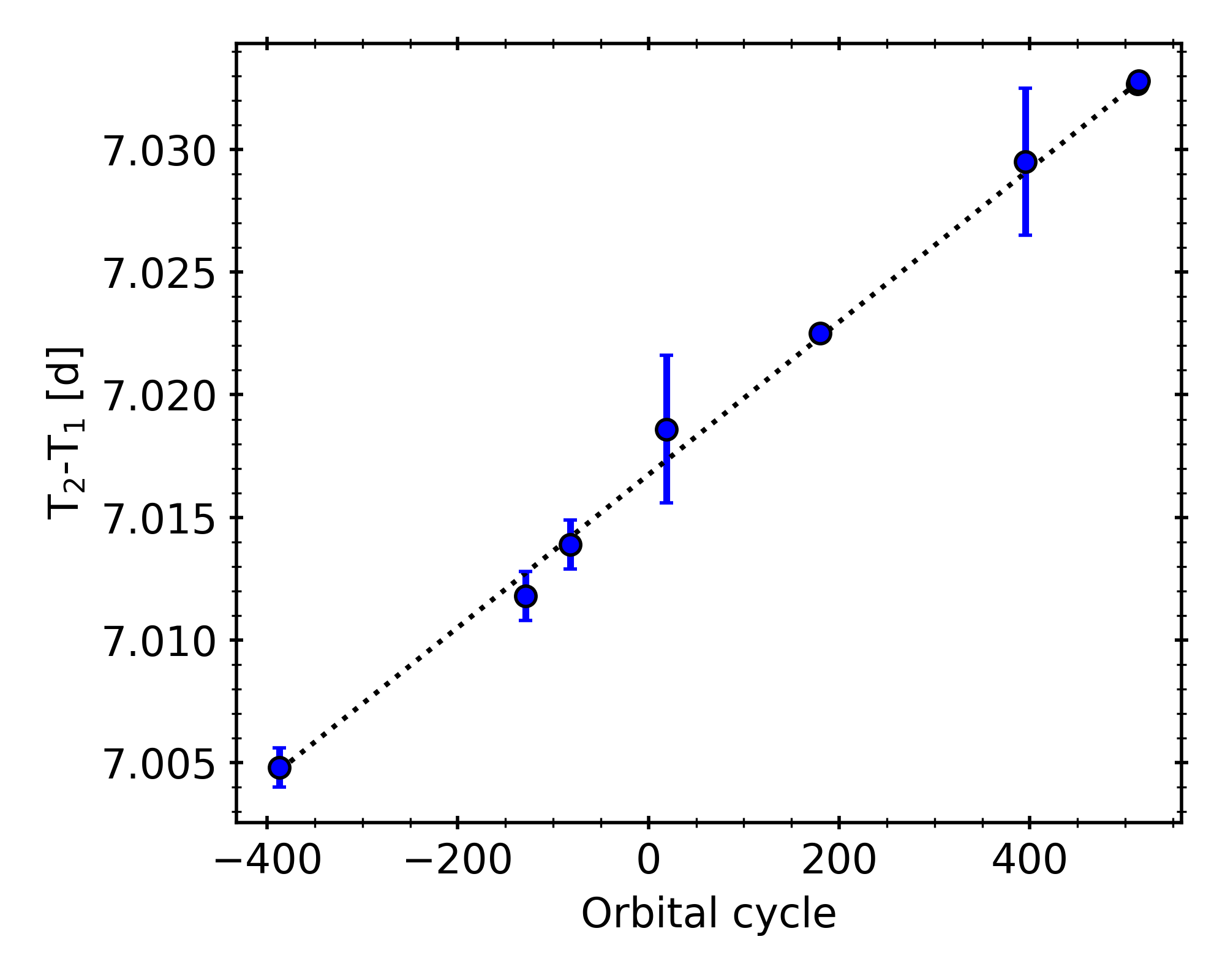}
\caption{$T_2-T_1$ as a function of the orbital cycle for V541\,Cyg.} 
          \label{fig:V541Cyg}%
\end{figure}

For our analysis, we included in Table\,\ref{tab:V541Cyg} the $T_2-T_1$ values derived from Table\,4 of \cite{Torres2017} which we complement with the values computed from the {\em TESS} data in Table\,\ref{tab:a1}. This was done using precise primary and secondary eclipses and the closest possible counterpart when calculating the timing differences. As mentioned by \cite{Torres2017}, some values had to be excluded because of their unusually large residuals. In Fig.\,\ref{fig:V541Cyg}, the variation of $T_2-T_1$ over time is shown and clearly indicates a well-defined increase of $(3.12 \pm 0.03) \times 10^{-5}$\,days\,cycle$^{-1}$. For the value corresponding to orbital cycle 19, the authors did not report any error bar and we assumed the same uncertainty as the largest one in Table\,\ref{tab:V541Cyg}, namely 0.003 days. As a simple consistency test, we checked that the observed slope yields $T_2-T_1$ = $6.9872 \pm 0.0003$ days at the time of the photographic light curve of \cite{Karpowicz1961}, while the actual measurement given by \cite{Torres2017} in their Table\,4, after a careful analysis of the original data, is $6.988\pm0.004$\,days. Both values are in very good agreement in spite of the extrapolation by about 550 orbital cycles. Furthermore, the predicted position of the secondary eclipse in phase at epoch 248 is $0.45798 \pm 0.00002$, in excellent agreement with the light-curve solution by \cite{Torres2017} in their Table\,5. Adopting $e = 0.4684 \pm 0.0014$ \citep{Torres2017}, we obtain an apsidal motion rate of $\dot{\omega} = 0.000352 \pm 0.000004$\,deg\,cycle$^{-1}$, more precise but in good agreement with the value of $\dot{\omega} = 0.000360 \pm 0.000012$\,deg\,cycle$^{-1}$ derived by \cite{Torres2017} using a completely different method, based on a simultaneous fit to photometric and spectroscopic data.

\begin{table}[t]
\centering
\caption{$T_2-T_1$ values used to compute the apsidal motion rate of V541\,Cyg.}
\label{tab:V541Cyg}
\begin{tabular}{rcl} 
\hline\hline
\noalign{\smallskip}
$N$ & $T_2-T_1$ [d]  & Ref.\\
\noalign{\smallskip}
\hline
\noalign{\smallskip}
$-$387  &   $7.0048\pm0.0008$       & Vol99  \\
$-$129  &   $7.0118\pm0.0010$       & DA \\
  $-$82 &   $7.0139\pm0.0010$       & LG  \\
   19   &   $7.0186\pm0.0030$       & Vol99  \\
 180    &   $7.0225\pm0.0003$       & WS  \\ 
 395    &   $7.0295\pm0.0030$       & Hub15  \\
\noalign{\smallskip}
\hline
\end{tabular}
\tablebib{Vol99: \cite{Volkov1999}; DA: \cite{Diethelm1992} and \cite{Agerer1994}; LG: \cite{Lacy1995} and \cite{Guinan1996}; WS: \cite{Wolf2010} and \cite{Smith2007}; Hub15: \cite{Hubscher2015a} and \cite{Hubscher2015b}.  
}
\end{table}

\subsection{V459\,Cas}

\cite{Lacy2004} determined the general properties of V459\,Cas and estimated the apsidal motion rate using all available minima at that time, obtaining a rather uncertain value of $\dot{\omega} = 0.0014 \pm 0.0011$\,deg\,cycle$^{-1}$. This could not be improved by \cite{Dariush2006} because of the narrow time-span of their reliable data. \cite{Torres2010} nevertheless quote an apsidal motion rate of $0.00057\pm0.00006$\,deg\,cycle$^{-1}$ as provided preliminarily to the authors by M. Wolf. This value was not confirmed later by \cite{Wolf2010}, who reported $0.00071\pm0.00008$\,deg\,cycle$^{-1}$, using again all available individual timings at that time.

\begin{figure}[t]
\centering
\includegraphics[width=\columnwidth]{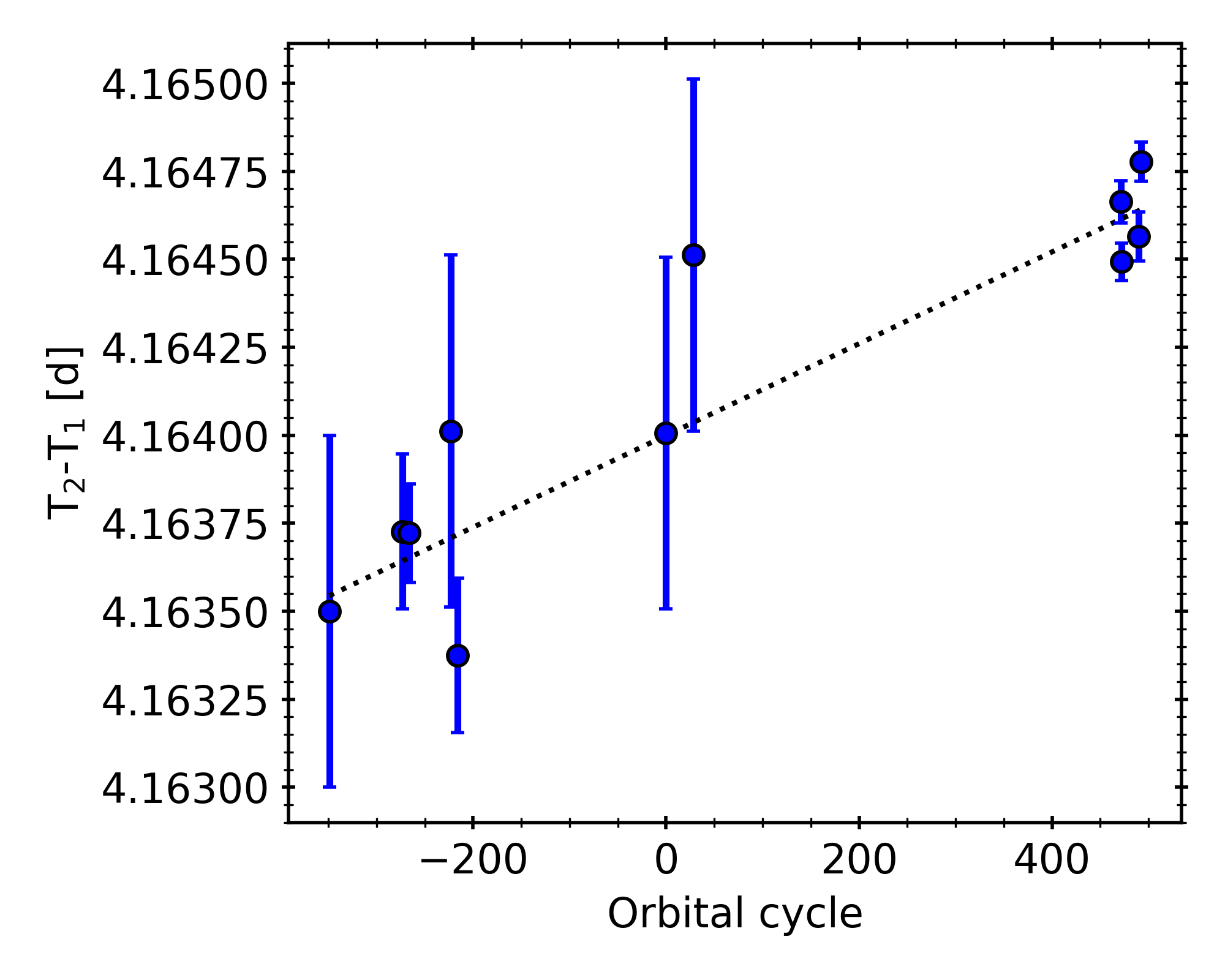}
\caption{$T_2-T_1$ as a function of the orbital cycle for V459\,Cas.} 
          \label{fig:V459Cas}%
\end{figure}

The low eccentricity of $0.0244\pm0.0004$ \citep{Lacy2004} of the system makes it difficult to determine the apsidal motion rate using the observed $T_2-T_1$ values. 
We searched in the list provided by \cite{Wolf2010}, in their Table A.1, and in Table 1 of \cite{Lacy2004} for the closest photoelectric pair of primary and secondary timings, which we list in Table\,\ref{tab:V459Cas}. The best linear fit to these values, together with the {\em TESS} $T_2-T_1$ measurements, is shown in Fig.\,\ref{fig:V459Cas}, and yields a slope of $(1.30\pm0.19)\times10^{-6}$\,deg\,cycle$^{-1}$, although with a large dispersion of the older timings due to the small variation in $T_2-T_1$ in the considered time-span. Using the orbital eccentricity of \cite{Lacy2004}, we obtain an apsidal motion rate of $\dot{\omega} = 0.00065 \pm 0.00010$\,deg\,cycle$^{-1}$, which agrees with the value given by \cite{Wolf2010}, although with a slightly larger uncertainty.

\begin{table}[t]
\centering
\caption{$T_2-T_1$ values used to compute the apsidal motion rate of V459\,Cas.}
\label{tab:V459Cas}
\begin{tabular}{rcl} 
\hline\hline
\noalign{\smallskip}
$N$ & $T_2-T_1$ [d]  & Ref.\\
\noalign{\smallskip}
\hline
\noalign{\smallskip}
$-349$ & $4.1634\pm0.0005$ & Lac01 \\
$-273$ & $4.16373\pm0.00022$ & Wol10 \\
$-266$ & $4.16372\pm0.00014$ & LN \\
$-223$ & $4.1640\pm0.0005$ & WH \\
$-216$ & $4.1634\pm0.0005$ & Lac04 \\
$0$ & $4.1640\pm0.0005$ & Wol10 \\
$28$ & $4.1645\pm0.0005$ & Wol10 \\
\noalign{\smallskip}
\hline
\end{tabular}
\tablebib{Lac01: \cite{Lacy2001IBVS}; Wol10: \cite{Wolf2010}; LN: \cite{Lacy2002IBVS} and \cite{Nelson2003}; WH: \cite{Wolf2010} and \cite{Hubscher2005}; Lac04: \cite{Lacy2004IBVS}.  
}
\end{table}

\subsection{V501\,Her}

The absolute parameters of V501\,Her are given in Table\,\ref{tab:systemsprop} as derived by \cite{Lacy2014} from their own light-curve and radial velocity measurements, from which no apsidal motion could be reported. The large relative radii indicate the evolved nature of the components and the difficulty in obtaining accurate times of eclipse due to their long duration. The {\em TESS} measurements given in Table\,\ref{tab:a1} present an internal dispersion that is larger than the estimated errors of the individual measurements. The weighted mean value is $T_2-T_1$ = $4.1208\pm 0.0004$ days, equivalent to a phase of the secondary eclipse of $0.47929 \pm 0.00005$ using the orbital period given by \cite{Lacy2014}. These authors give a secondary eclipse phase of $0.4791 \pm 0.0002$, implying that no significant variation is observed. From the most complete and precise light curve in \cite{Lacy2014} and using the elements in their Table 4, we estimated the phase of the secondary eclipse at $0.47915\pm0.00005$, compatible with but more precise than the one given by their times of eclipse, and equivalent with a $T_2-T_1$ of $4.1196\pm0.0004$\,d. Comparing this value with that from the {\em TESS} observations, we measure a change in $T_2-T_1$ of $0.0012 \pm 0.0006$ days over 354 orbital cycles, yielding a slope of $(3.4\pm1.7)\times 10^{-6}$\,deg\,cycle$^{-1}$. This shows the presence of apsidal motion in V501\,Her but with a poorly determined rate. The corresponding apsidal motion rate is $\dot{\omega} = 0.00041 \pm 0.00020$\,deg\,cycle$^{-1}$.

\subsection{KW\,Hya}

The only accurate masses and radii for this system date back to \cite{Andersen1984}, resulting from the analysis of both light and radial velocity curves. Such results were confirmed more recently by \cite{Gallenne2019} with astrometric observations, and we adopted those values as listed in Table\,\ref{tab:systemsprop}. Apsidal motion was not reported for KW\,Hya, despite the large time-span between the two orbital studies, because of the uncertainties involved in the determination of the argument of periastron. 

The weighted mean of the {\em TESS} measurements in Table\,\ref{tab:a1} provides an accurate value of the $T_2-T_1$ difference of $3.54987 \pm 0.00004$ days. This can be compared with the corresponding timing difference measured by \cite{Andersen1984}, 1740 orbital cycles before, of $3.5454 \pm 0.0007$ days. No other value of sufficient precision could be found in the literature. These two available measurements yield a slope of $2.6 \pm 0.4 \times 10^{-6}$\,days\,cycle$^{-1}$. Adopting the eccentricity given by \cite{Gallenne2019}, $e = 0.094 \pm 0.004$, gives an apsidal motion rate of $\dot{\omega} = 0.00045 \pm 0.00007$\,days\,cycle$^{-1}$.

\subsection{V501\,Mon}

A detailed analysis of the light and radial velocity curves of V501\,Mon was carried out by \cite{Torres2015} and the corresponding absolute parameters are given in Table\,\ref{tab:systemsprop}. In their study, an analysis of the available times of eclipse at that time was carried out together with the radial velocities in order to determine the orbital elements and the possible variation in the argument of periastron  simultaneously. The authors obtained an estimate of the apsidal motion rate of $0.00045 \pm 0.00024$\,deg\,cycle$^{-1}$, which we now try to improve by combining the {\em TESS} measurements given in Table\,\ref{tab:a1} with the data used by \cite{Torres2015}. The weighted mean of the {\em TESS} $T_2-T_1$ values gives a separation between primary and secondary eclipses of $3.14676 \pm 0.00020$, which does not show significant variation with respect to the solution by \cite{Torres2015}. 

\begin{figure}[t]
\centering
\includegraphics[width=\columnwidth]{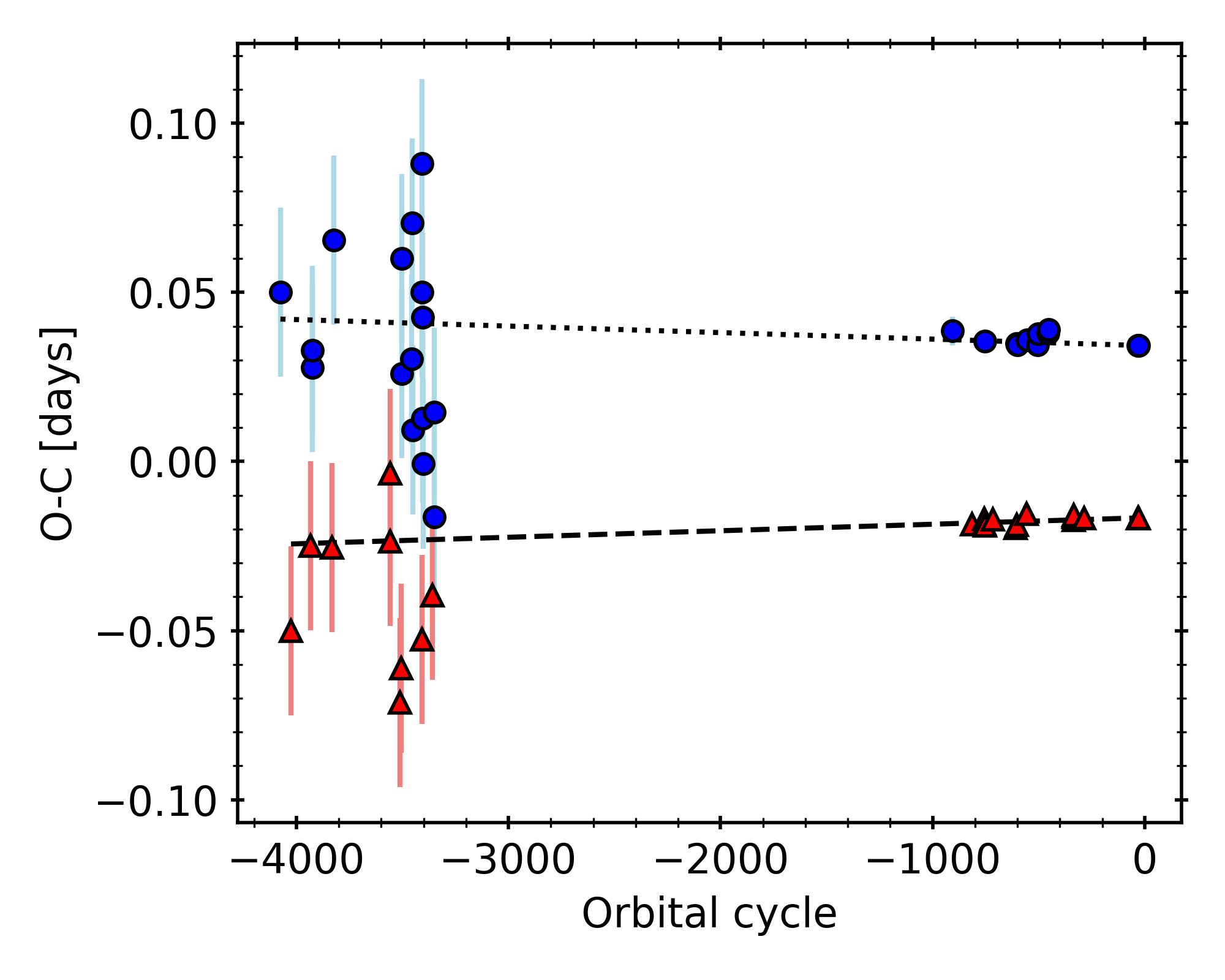}
\caption{Same as Fig.\,\ref{fig:ALDor} for V501\,Mon.} 
          \label{fig:V501Mon}%
\end{figure}

We then fitted all the individual timings corrected to BJD in order to identify a possible difference between the linear periods of the primary and secondary eclipses. Using all the times of eclipse available in Table 1 of \cite{Torres2015} (adopting also the same scaling for the photoelectric timings) and the {\em TESS} individual eclipses, we computed the periods resulting from primary and secondary eclipses, and obtained a difference of $\Delta P = 3.85 \pm 0.89 \times 10^{-6}$\,days\,cycle$^{-1}$, which corresponds to the left-hand side term in Eq.\,(\ref{eqt2t1}) \citep{Gimenez1995}. With the value of the eccentricity $e = 0.1339 \pm 0.0006$, this $\Delta P$ value yields an apsidal motion rate of $\dot{\omega} = 0.00046 \pm 0.00011$\,deg\,cycle$^{-1}$ , which is in excellent agreement with the value published by \cite{Torres2015} but with a reduced uncertainty. Figure\,\ref{fig:V501Mon} displays the linear best fit to primary and secondary individual timings. The data points on the left, showing a large dispersion, correspond to the photographic measurements in Table 1 of \cite{Torres2015}, for which we assumed uncertainties of 0.025\,d.

\subsection{GG\,Ori}

The general properties of GG\,Ori were determined by \cite{Torres2000}, as listed in Table\,\ref{tab:systemsprop}, together with an apsidal motion rate of $\dot{\omega} = 0.00061 \pm 0.00025$\,deg\,cycle$^{-1}$. \cite{Torres2000} followed a method based on the combination of eclipse timings with radial velocities, the same used in their analysis of V501\,Mon mentioned above. Later, \cite{Wolf2010} revised the analysis of the individual eclipse timings available at the time and obtained a more precise $\dot{\omega} = 0.00057 \pm 0.00006$\,deg\,cycle$^{-1}$, with an orbital eccentricity of $e = 0.220 \pm 0.001$.

\begin{figure}[t]
\centering
\includegraphics[width=\columnwidth]{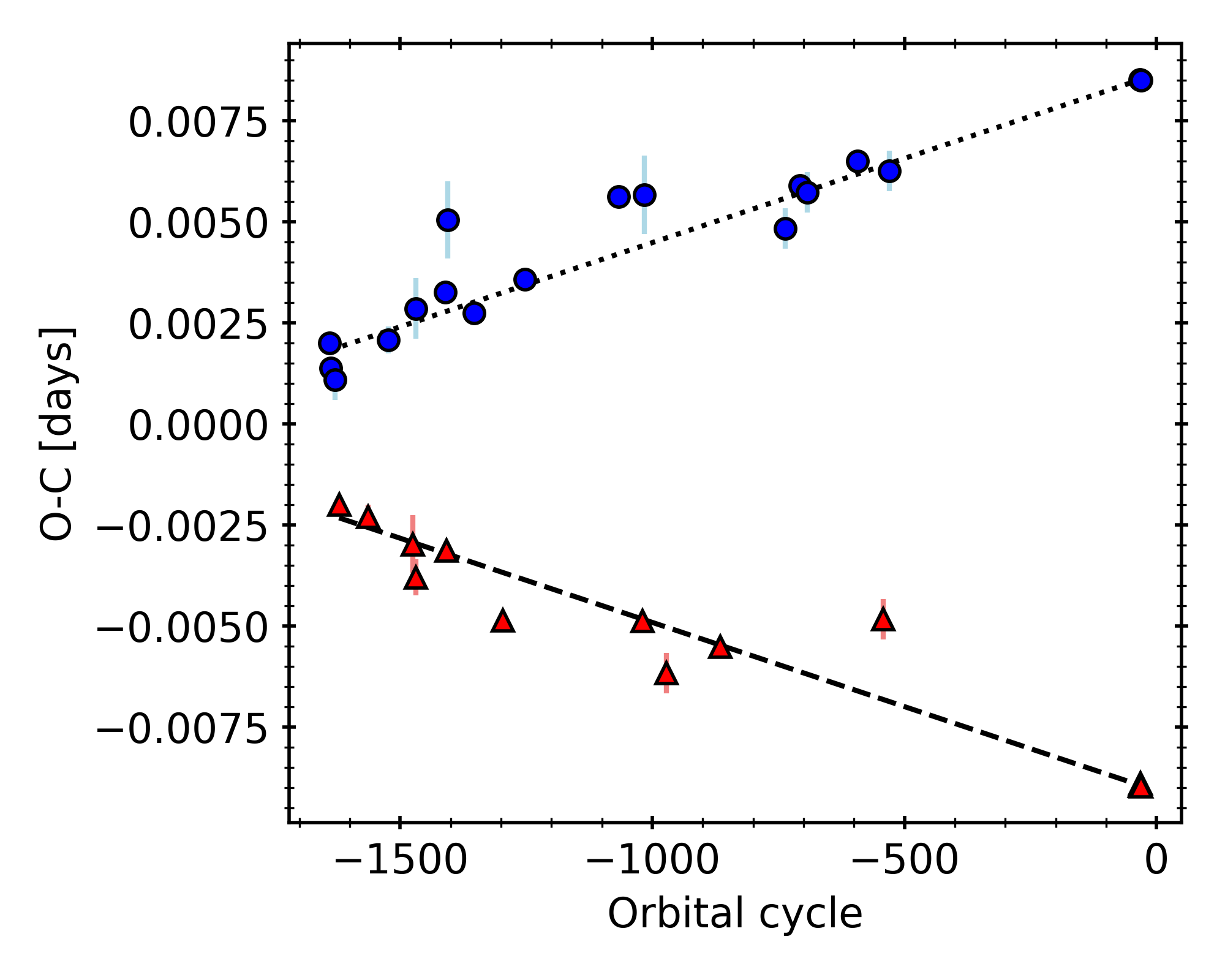}
\caption{Same as Fig.\,\ref{fig:ALDor} but for GG\,Ori.} 
          \label{fig:GGOri}%
\end{figure}

In order to improve the apsidal motion rate determination combining early timings with {\em TESS}, we compiled $T_2-T_1$ measurements from the best minima given by \cite{Wolf2010} acquired within less than ten orbital cycles, but only three pairs of primary and secondary eclipses met this criterion. We derived $\dot{\omega} = 0.00060 \pm 0.00003$\,deg\,cycle$^{-1}$. We then performed linear fits to primary and secondary individual timings, respectively, obtaining the linear period for primary and secondary eclipses and obtained a difference of $\Delta P = -8.33 \pm 0.16 \times 10^{-6}$\,days\,cycle$^{-1}$ using only those eclipses with weight ten in \cite{Wolf2010}. Furthermore, using the eccentricity obtained by \cite{Torres2000}, $e = 0.2218 \pm 0.0022$, we obtain an apsidal motion rate of $0.00061 \pm 0.00003$\,deg\,cycle$^{-1}$ , in good agreement with the previous determinations albeit much more precise. The ephemeris curve of the individual primary and secondary eclipse timings is shown in Fig.\,\ref{fig:GGOri}.

\subsection{EY\,Cep}

This binary system was studied by \cite{Lacy2006}, who obtained the general properties given in Table\,\ref{tab:systemsprop}. Given the short time-span of their photometric and spectroscopic observations, the authors were not able to detect any indication of apsidal motion. The new {\em TESS} data listed in Table\,\ref{tab:a1} include two sectors and the $T_2-T_1$ variation already indicates the presence of apsidal motion, though the modest time-span does not allow a precise rate determination. We further used the eclipses given by \cite{Lacy2006} to obtain the time differences in Table\,\ref{tab:EYCep}.

\begin{table}[t]
\centering
\caption{$T_2-T_1$ values used to compute the apsidal motion rate of EY\,Cep.}
\label{tab:EYCep}
\begin{tabular}{rcl} 
\hline\hline
\noalign{\smallskip}
$N$ & $T_2-T_1$ [d]  & Ref.\\
\noalign{\smallskip}
\hline
\noalign{\smallskip}
$-817$ &  $3.1565\pm0.0003$ & Lac06  \\
$-815$ &  $3.1564\pm0.0004$ & Lac06 \\
$-781$ &  $3.1557\pm0.0005$ & Lac06 \\
\noalign{\smallskip}
\hline
\end{tabular}
\tablebib{
Lac06: \cite{Lacy2006}. 
}
\end{table}

\begin{figure}[t]
\centering
\includegraphics[width=\columnwidth]{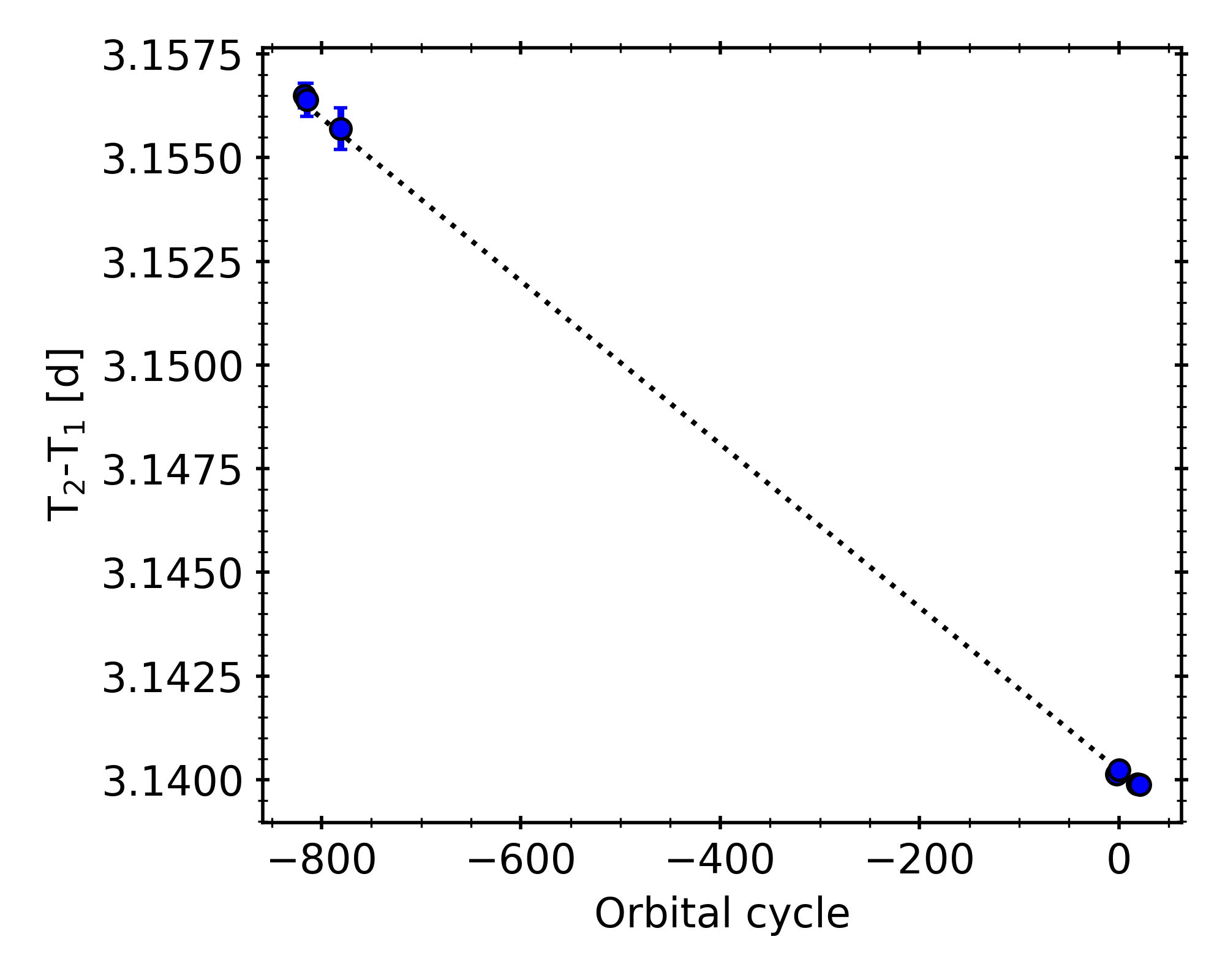}
\caption{$T_2-T_1$ as a function of the orbital cycle for EY\,Cep.} 
          \label{fig:EYCep}%
\end{figure}

The graphical representation of the variations is shown in Fig.\,\ref{fig:EYCep} and the linear least-squares fit yields a well-defined slope of ($-1.97 \pm 0.06) \times 10^{-5}$\,days\,cycle$^{-1}$. Using the eccentricity given by \cite{Lacy2006}, $e=0.4429\pm0.0014$, this yields an apsidal motion rate of $\dot{\omega} = 0.000507 \pm 0.000016$\,deg\,cycle$^{-1}$, which is very precise thanks to the accurate {\em TESS} data, the high orbital eccentricity, and a time coverage of more than 800 orbital cycles. 

\section{Comparison between theory and observations}\label{sec:results}

The values of $\dot{\omega}_{\rm obs}$ obtained from the apsidal motion analysis using data from {\em TESS} and archival times of minima are listed in Table\,\ref{tab:results}. Of the nine systems studied for which we were able to perform apsidal motion rate determinations, five are reported for the first time. The precision of the determinations varies widely but they are well above the 3$\sigma$ threshold, except for the case of V501\,Her, which has an apsidal motion rate determination with a significance of only 2$\sigma$. Table\,\ref{tab:results} provides the values of $\dot{\omega}_{\rm rel}$ and $\dot{\omega}_{\rm cl}$ for the systems with detected apsidal motion, computed using Eqs.\,(\ref{eqrel2}) and (\ref{eqclas1}). Calculating the classical term requires employing two additional parameters for each component, namely the internal structure constant, $k_2$, and the rotational velocity.

The values of $k_2$ given in Table\,\ref{tab:results} were specially calculated by A. Claret using theoretical models based on the Modules for Experiments in Stellar Astrophysics package \citep[MESA;][]{Paxton2011,Paxton2013,Paxton2015} and following the methodology described in the series of papers by \cite{Claret2017b,Claret2018b,Claret2019b}. A coarse-grid search was performed over evolutionary tracks calculated for the measured masses of each component, allowing the convective core overshooting parameter ($f_{ov}$) and the mixing length parameter ($\alpha_{\rm MLT}$) to vary freely, with a variable metallicity ($Z$) common to both components. The $\log k_2$ values were determined from the best match between the grid of evolutionary tracks and the observed masses, radii, and effective temperatures. The theoretical internal structure constants, $k_2$, for the models fitting the observed parameters, were integrated using the differential equations of Radau as given by Eqs. (1) to (3) in \cite{Claret2010b}.

The other parameter used to estimate $\dot{\omega}_{\rm cl}$ is the rotational velocity. The values listed in Table\,\ref{tab:otherprop} were taken from the same spectroscopic analyses in the literature from which we adopted the general properties in Table\,\ref{tab:systemsprop}. Nevertheless, in the case of AL\,Dor, no rotational velocity was reported for the component stars and we used the predicted values under the assumption of pseudo-synchronisation, as described by \cite{Hut1981}. 

\begin{figure}[t]
\centering
\includegraphics[width=\columnwidth]{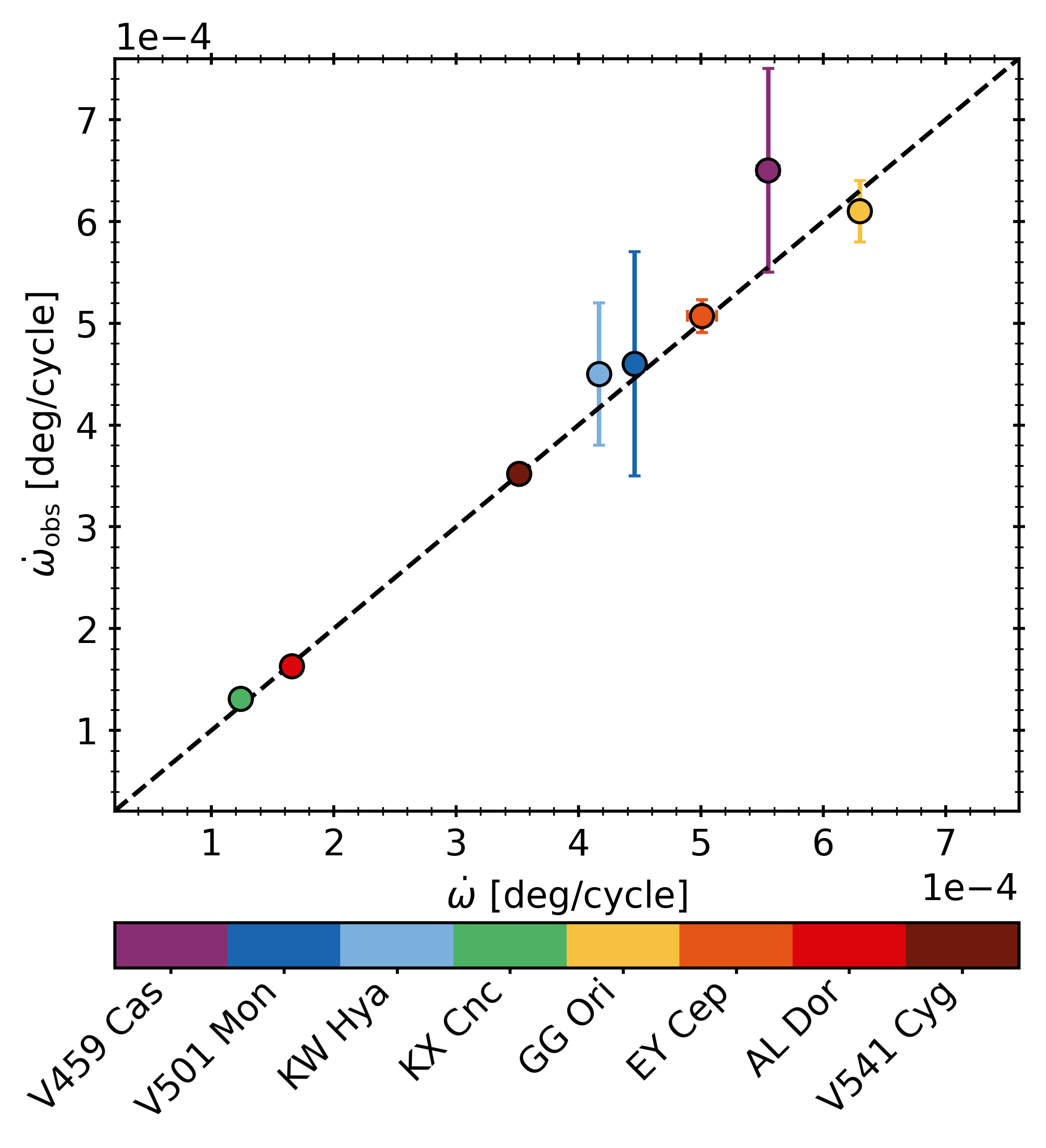}
\caption{Comparison between total computed and total observed apsidal motion rates. The black dashed line indicates the 1:1 relation.} 
          \label{fig:obspred}%
\end{figure}

All the parameters needed to apply Eqs.\,(\ref{eqrel2}) and (\ref{eqclas1}) are listed in Tables\,\ref{tab:systemsprop} and \ref{tab:otherprop}. The errors of the input parameters were propagated to obtain the uncertainty in $\dot{\omega}_{\rm cl}$ and $\dot{\omega}_{\rm rel}$. The uncertainties of the stellar rotation and $k_2$ dominate the error budget in the classical term, while the uncertainties in the component masses are the main source of error in the GR term. In Figure\,\ref{fig:obspred} we compare the observed apsidal motion rates with the total calculated theoretical value $(\dot{\omega} = \dot{\omega}_{\rm cl} + \dot{\omega}_{\rm rel})$, as given in Table\,\ref{tab:results}. We excluded V501\,Her from Figure\,\ref{fig:obspred} given the poor significance of the apsidal motion determination (below 2$\sigma$), although the observed value agrees with theory within the uncertainties. 

\begin{table*}[t]
\centering
\caption{$\log k_2$ values and apsidal motion rates, both observed and theoretically predicted, for the eclipsing binaries with apsidal motion measurement.}
\label{tab:results}
\begin{tabular}{llllllllll} 
\hline\hline
\noalign{\smallskip}
\multirow{2}{*}{System} & \multirow{2}{*}{$\log k_{2,1}$} & \multirow{2}{*}{$\log k_{2,2}$} &  \multicolumn{1}{c}{$\dot{\omega}_{\rm obs}$}  &  \multicolumn{1}{c}{$\dot{\omega}$}  &  \multicolumn{1}{c}{$\dot{\omega}_{\rm cl}$}  &  \multicolumn{1}{c}{$\dot{\omega}_{\rm rel}$} &  \multicolumn{1}{c}{$\dot{\omega}_{\rm rel,mea}$} &   \multicolumn{1}{c}{\multirow{2}{*}{$\dot{\omega}_{\rm rel,mea}$/ $\dot{\omega}_{\rm rel}$}} \\
  &  & &  \multicolumn{1}{c}{[deg\,cycle$^{-1}$]}  &   \multicolumn{1}{c}{[deg\,cycle$^{-1}$]} &   \multicolumn{1}{c}{[deg\,cycle$^{-1}$]}  &   \multicolumn{1}{c}{[deg\,cycle$^{-1}$]} &   \multicolumn{1}{c}{[deg\,cycle$^{-1}$]} & \\
\noalign{\smallskip}
\hline
\noalign{\smallskip}
KX\,Cnc & $-$1.88(4) & $-$1.88(4) & 0.000131(10) &      0.0001241(2) &  0.0000029(1) &       0.00012127(16) &        0.000128(10) & 1.06(8) \\
AL\,Dor & $-$1.95(3) & $-$1.95(4) & 0.000163(6) &       0.0001659(4) &  0.0000075(4) &       0.00015838(3) & 0.000156(6) & 0.98(4) \\
V541\,Cyg & $-$2.37(3) & $-$2.34(3) & 0.000352(4) &     0.0003516(21) & 0.0000392(17) &       0.0003124(12) & 0.000313(4) & 1.001(14) \\
V459\,Cas & $-$2.48(3) & $-$2.47(3) & 0.00065(10) &     0.000555(9) &   0.000226(9) & 0.0003297(23) & 0.00043(10) & 1.3(3) \\
V501\,Her & $-$2.13(3) & $-$1.96(4) & 0.00041(20) &     0.000521(13) &  0.000281(13) &       0.0002400(3) &  0.00013(20) & 0.5(8) \\
KW\,Hya & $-$2.50(3) & $-$2.45(10) & 0.00045(7) &       0.000417(7) &   0.000096(7) &       0.0003211(19) & 0.00035(7) &1.10(22) \\
V501\,Mon & $-$2.54(3) & $-$2.52(3) & 0.00046(11) &     0.000446(8) &   0.000124(8) & 0.0003219(3) & 0.00034(11) & 1.0(3) \\
GG\,Ori & $-$2.336(20) & $-$2.337(20) & 0.00061(3) &    0.000630(9) &   0.000176(9) &       0.0004541(16) &         0.00043(3) &0.96(7) \\
EY\,Cep & $-$2.38(5) & $-$2.40(5) & 0.000507(16) &      0.000501(12) &  0.000146(12)  &      0.0003549(14) & 0.000361(20) & 1.02(6) \\
\noalign{\smallskip}
\hline
\end{tabular}
\end{table*}

The comparison of the observed apsidal motion rates with those calculated using Eqs. (\ref{eqrel2}) and (\ref{eqclas1}), and the parameters in Tables\,\ref{tab:systemsprop} and \ref{tab:otherprop}, show good agreement and no systematic deviations within their uncertainties. An equivalent approach to test GR effects was employed before by \cite{Torres2010}, who also considered systems with accurate masses and radii and limited the comparison between observed and theoretical apsidal motion rates to those having a relative contribution of the GR term of at least 40\%. Figure\,\ref{fig:obspred} therefore complements Fig.\,11 in \cite{Torres2010}, including additional systems with a relative contribution of the GR term above 60\%. The values used by \cite{Gimenez1985,Gimenez2007}  for the system's general properties were of insufficient precision to allow for a meaningful and unbiased test of the GR term. \cite{Claret1993} searched for systematic deviations in the comparison between observed and theoretical values of $\log k_2$ to falsify the predictions of the \cite{Moffat1986} theory of gravitation. \cite{Claret1997} and \cite{Wolf2010} studied the complementary problem, where the GR term was theoretically estimated and subtracted from the observed value. The resulting classical term was then compared with model predictions of the $k_2$ parameter. 

\begin{figure}[t]
\centering
\includegraphics[width=\columnwidth]{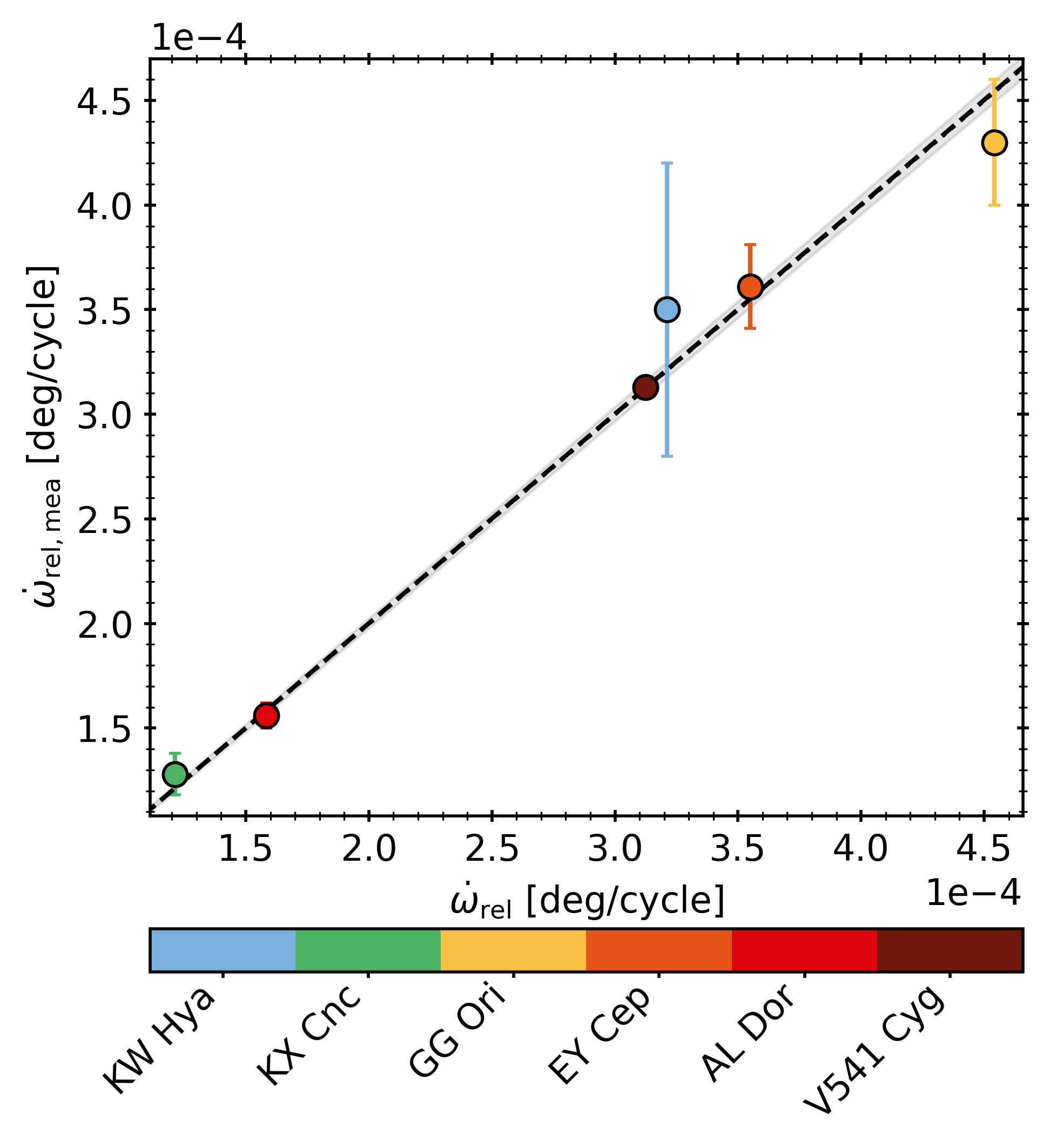}
\caption{Measured GR apsidal motion as a function of the theoretically computed GR apsidal motion. The black dashed line and grey shadow area correspond to the best fit to Eq.\,(\ref{eqall}), assuming $\alpha_i=\zeta_2=0$, and its 1$\sigma$ uncertainty. The colour code is chosen such that redder colours correspond to systems with the smaller relative error, and therefore are dominating the fit.} 
          \label{fig:relobs}%
\end{figure}

\section{A test of gravitational theories} \label{sec:test}

\begin{figure}[!t]
\centering
\includegraphics[width=\columnwidth]{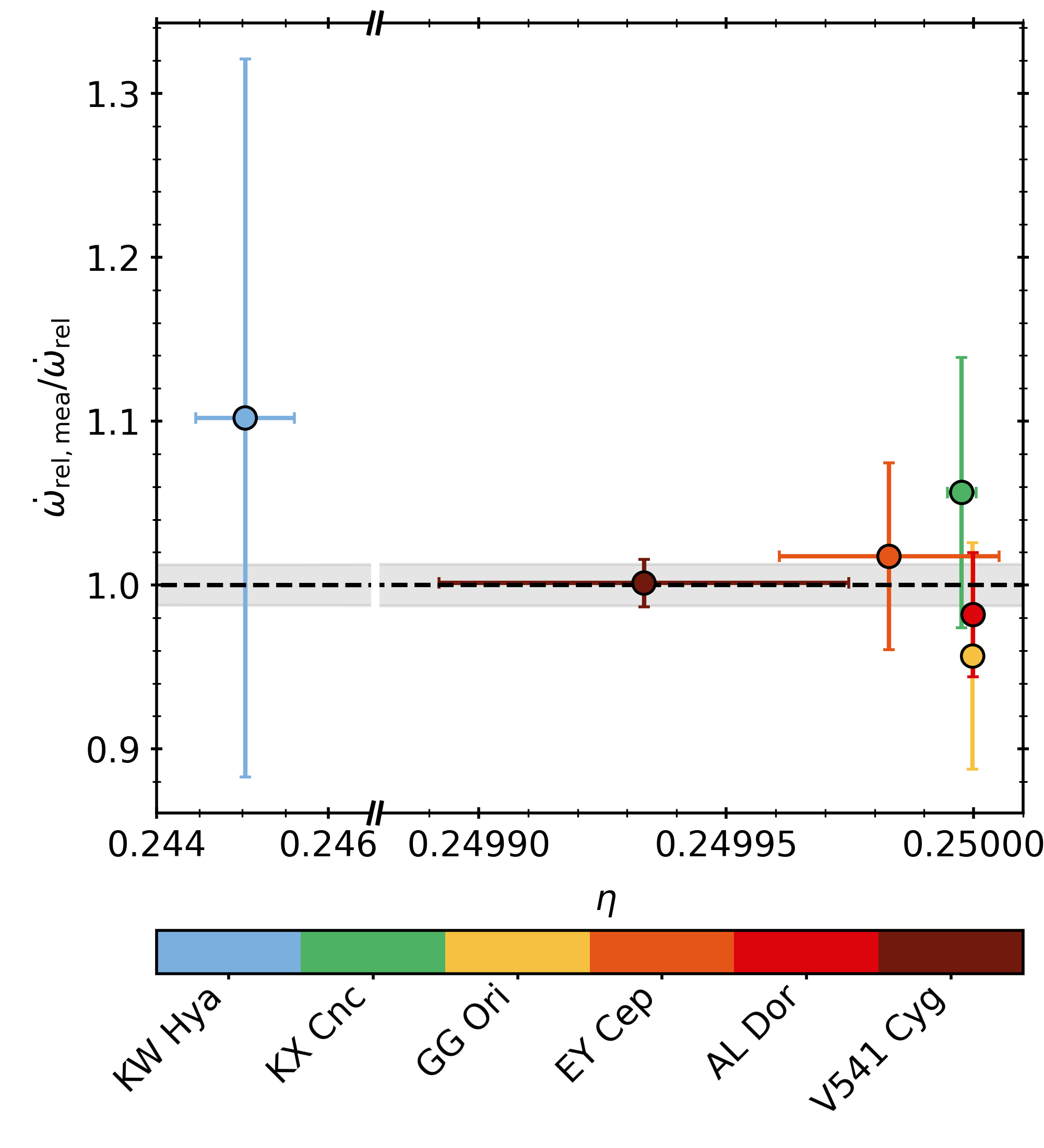}
\caption{Ratio of measured and predicted GR apsidal motion rates as a function of the dimensionless reduced mass $\eta$. The black dashed line and grey shadow area correspond to the best fit of Eq.\,(\ref{eqall}), assuming $\gamma=\beta=1$, and its 1-$\sigma$ uncertainty. We highlight the broken x-axis and the different scales. The colour code is chosen such that redder colours correspond to systems with the smaller relative error, and therefore are dominating the fit.} 
          \label{fig:eta}%
\end{figure}

For a more in-depth analysis, the two apsidal motion components should be analysed separately. As explained above, we carefully selected the sample to ensure a maximum contribution of the GR term, so that, after subtraction of the (model-calculated) quadrupole Newtonian contribution, a comparison with gravitational theories could be carried out. \cite{DeLaurentis2012} attempted a test of gravitational theories using a similar method but their eclipsing binary sample was not appropriate, with over half of their sample having a GR term contribution below 10\%. Their results were therefore inconclusive.

Thanks to the improved precision in the apsidal motion determinations that we have attained, we can perform a comparison of the predicted theoretical relativistic apsidal motion rate with the difference between the observed values and the computed classical terms, as listed in Table\,\ref{tab:results}. For this procedure to succeed, the uncertainty in the classical contribution needs to be well below this latter difference and therefore this limits the comparison to systems with a small non-relativistic contribution to the observed apsidal motion. In Table\,\ref{tab:results} we also include the measured GR rate calculated as $\dot{\omega}_{\rm rel,mea}=\dot{\omega}_{\rm obs}-\dot{\omega}_{\rm cl}$, where $\dot{\omega}_{\rm obs}$ is the observed apsidal motion from precise minima timings derived in this work. Given the stringent requirements on the system characterisation (to ensure good estimates of the internal structure parameters and classical apsidal motion contribution), we excluded the cases of V459\,Cas and V501\,Mon in spite of their good agreement shown in Figure\,\ref{fig:obspred}.

Figure\,\ref{fig:relobs} shows the comparison between predicted $\dot{\omega}_{\rm rel}$, according to Eq.\,(\ref{eqrel2}), and $\dot{\omega}_{\rm rel,mea}$. The measured values for all systems are compatible with the GR predictions within their errors. A more general form of Eq.\,(\ref{eqrel2}), which also allows to test other possible gravitation theories in addition to GR is given by equation 66 of \cite{Will2014}. This uses the parametrised post-Newtonian (PPN) formalism, with several parameters whose values depend on the gravitation theory chosen:
\begin{equation}
    \dot{\omega}_{\rm rel,mea}=\dot{\omega}_{\rm rel}\left(\frac{1}{3}(2+2\gamma-\beta)+\frac{1}{6}(2\alpha_1-\alpha_2+\alpha_3+2\zeta_2)\eta \right).
    \label{eqall}
\end{equation}
Here, $\dot{\omega}_{\rm rel}$ is the value of the apsidal motion predicted by GR, which is given by Eq.\,(\ref{eqrel2}), $\eta \equiv M_1 M_2/(M_1+M_2)^2$ is the dimensionless reduced mass, and $\gamma$, $\beta$, $\zeta_2$, and $\alpha_i$ are the PPN parameters. In any fully conservative theory of gravity, $\alpha_i = \zeta_2 \equiv 0$, while for GR, $\gamma = \beta \equiv1$ and $\alpha_i = \zeta_2 \equiv 0$, recovering the expression in Eq.\,(\ref{eqrel2}). The PPN formalism is appropriate for weak gravitational fields and slow motions, such as the conditions in stellar eclipsing binary systems \citep[see][for more details]{Will2014}.

Therefore, a test of GR can be performed by checking that the measured and theoretical GR apsidal motion rates are compatible with values of $\alpha$ and $\beta$ predicted by the theory, assuming a fully conservative model (i.e. $\alpha_i = \zeta_2 \equiv 0$). The black dashed line shown in Fig.\,\ref{fig:relobs} corresponds to the best fit to the measured and predicted values of the GR apsidal motion, assuming only a varying slope. This slope, as deduced from Eq.\,(\ref{eqall}), corresponds to $A \equiv (2+2\gamma-\beta)/3$. We derive a value of $A = 1.002\pm0.012$, which is fully compatible with the value predicted by GR of 1. 
It is clear from Fig.\,\ref{fig:relobs} that the stronger constraints are set by the systems with the smaller relative uncertainties, that is, AL\,Dor and V541\,Cyg. However, it should be noted that the fit is still compatible with GR when removing these two targets, obtaining $A=1.015\pm0.036$. 

From our determination of $A$, the validity of alternative gravitational theories in the PPN formalism can be assessed. For example, constraints can be put on the coupling constant $\omega_{BD}$ of the Brans-Dicke theory \citep{Estabrook1969}, which takes $A= (4+3\omega_{BD})/(6+3\omega_{BD})$. Furthermore, our measurements can also be used to place bounds on the standard individual PPN parameters. Adopting the limit of $(\gamma-1)$ from the Shapiro time-delay measurements using the Cassini spacecraft, of $2.3\times 10^{-5}$ \citep{Bertotti2003}, our measurement yields a constraint on $\beta$ given by $(\beta-1)=-0.005\pm0.035$. 
Equivalently, adopting the limit of $(\beta-1)$ from the perihelion shift of Mercury, of $8\times 10^{-5}$ \citep{Verma2014}, we find a constraint to $\gamma$ of $(\gamma-1)=0.002\pm0.017$.

A test of non-conservative models can be carried out by measuring the value of $B\equiv(2\alpha_1-\alpha_2+\alpha_3+2\zeta_2)/6$ in Eq.\,(\ref{eqall}). Assuming now $\gamma=\beta=1$, we can determine the value of $B$ by fitting the relation $\dot{\omega}_{\rm rel,mea}/\dot{\omega}_{\rm rel}=1+B\,\eta$. Such an assumption is justified by the measurements of the perihelion shift of Mercury, which correspond to a very small value of $\eta$, and determine $A=1$ at high significance. In the last column of Table\,\ref{tab:results} we list the ratio between the measured and predicted GR apsidal motion rates, which we plot  as a function of $\eta$ in Fig.\,\ref{fig:eta}. The black dashed line and the grey shadow region represent the best fit of $B$ and the 1$\sigma$ uncertainty. We obtain $B=0.01\pm0.05$, 
which again is fully consistent with the predicted null value from GR. Only one system, KW\,Hya, has a value of $\eta$ that is significantly different from the bulk of the sample at $\eta=0.250$. This clustering of measurements at a small range in $\eta$ limits the effectiveness of the sample at placing strong constraints on the slope $B$. For this reason, we cannot perform a fit varying simultaneously $A$ and $B$ from Eq.\,(\ref{eqall}). This would be enabled by considering systems with very unequal component masses, which are difficult to discover and measure with high precision.

\begin{figure}[t]
\centering
\includegraphics[width=\columnwidth]{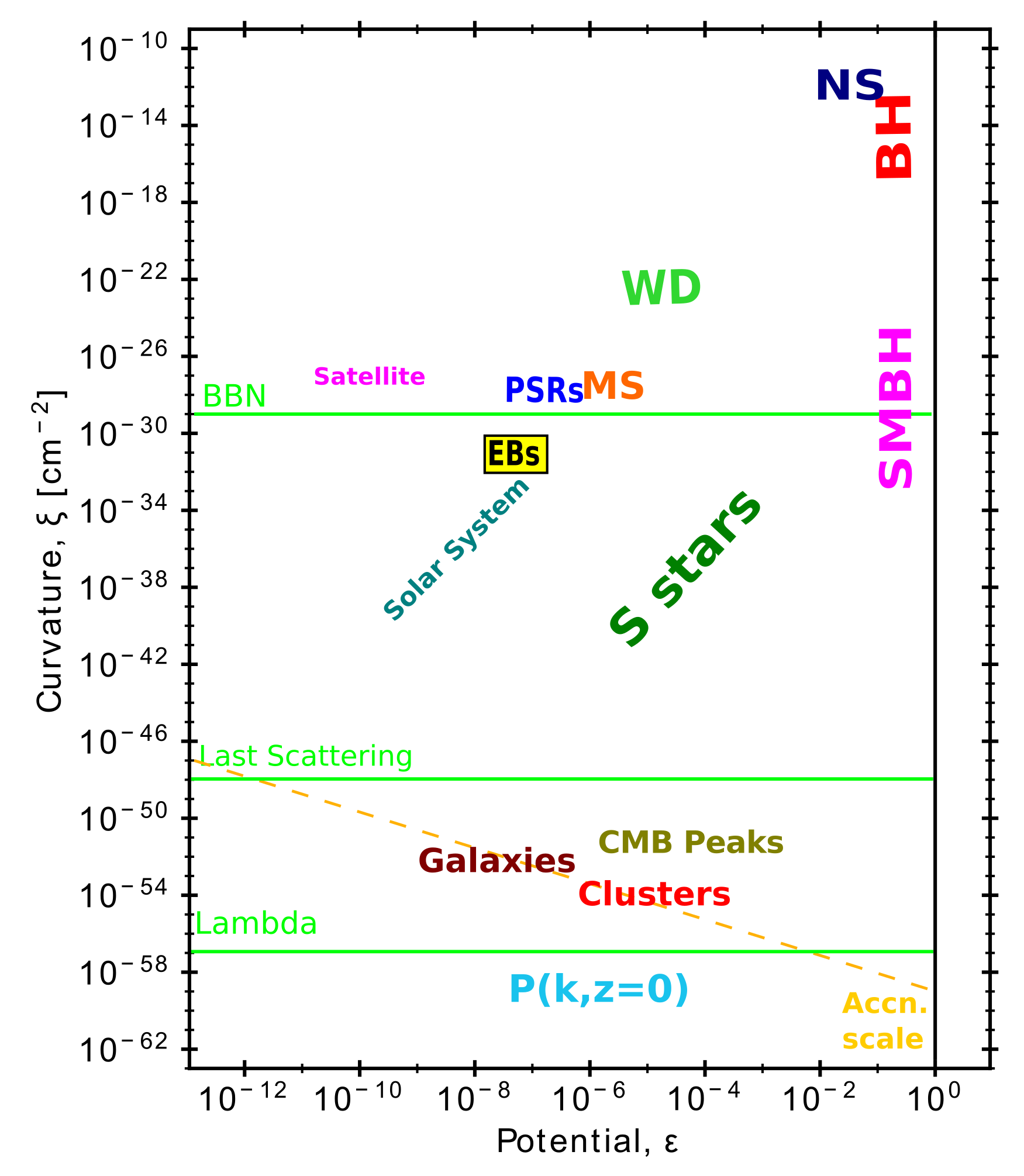}
\caption{Adaptation of Fig. 1 of \cite{Baker2015} showing the parameter space for gravitational fields, including the regime probed by different astrophysical and cosmological systems. We add the regime tested by the eclipsing binaries in this work as a yellow square. See \cite{Baker2015} for more information about the labels.} 
          \label{fig:potcurv}%
\end{figure}

Our analysis provides weaker constraints to PPN parameters compared to for example those from the perhielion shift of Mercury \citep{Verma2014}, the Shapiro time-delay measured by the Cassini mission \citep{Bertotti2003}, or the spin precession of millisecond pulsars \citep{Shao2013}. Nevertheless, the results obtained from the eclipsing binaries analysed in this work perform a test to gravity in a yet unexplored regime of the potential-curvature diagram. Figure\,\ref{fig:potcurv} reproduces this latter diagram from \cite{Baker2015} where we add the systems studied here. Stellar eclipsing binaries are one of the many probes of gravity in different environments, in this case covering the range between Solar System planets and binary pulsars.

\section{Conclusions} \label{sec:conclusions}

The goal of the present study is to employ precise and broad photometric coverage provided by the {\em TESS} mission to determine precise eclipse timings of a number of binaries and determine their apsidal motion rates. We carefully selected a sample of eccentric eclipsing binaries with accurately determined general properties such as masses and radii, and relatively long orbital periods to ensure a dominant contribution of the GR term over the classical term to the apsidal motion rate. The continuous monitoring and the excellent precision of the {\em TESS} new times of minima allowed us to derive determinations of $T_2-T_1$ and therefore investigate possible variations. From our analysis we were able to detect apsidal motion in 9 out of the 15 systems included in our sample. In some cases, we improved previous determinations, but for 5 of the systems (KX\,Cnc, AL\,Dor, V501\,Her, KW\,Hya, and EY\,Cep) we present the first measurement of apsidal motion. Furthermore, we were able to determine apsidal motion in one system, AL\,Dor, using data from {\em TESS} alone, which cover a very small fraction of the apsidal motion orbital cycle, thus highlighting the excellent performance of the mission for this kind of study. 

A comparison between observed apsidal motion rates and theoretically predicted values yields excellent agreement. The good precision of the observed apsidal motion rates, the accurate stellar properties employed, and the relatively modest contribution from the classical term made it possible to calculate reliable estimates of the observed GR term. In some cases, when the classical contribution is especially insignificant, the GR apsidal motion could be measured with a relative precision approaching 1 per cent. This allows a test of GR for the first time using this method. Our results strongly favour the predictions of GR, with no deviations observed at the level of $10^{-2}$. Furthermore, we set constraints on two of the PPN parameters. These latter are not quite as stringent as those resulting from other methods \citep[which can be three orders of magnitude more constraining;][]{Will2014} yet they probe a regime of gravitational forces and potentials that has not been explored before \citep{Baker2015}. Given the excellent agreement with GR, the parameter space for alternative theories continues to narrow. 

The results we present are mildly model-dependent as they may be affected by deviations of the $k_2$ values. However, because of the small contribution of the classical term, variations of the $k_2$ values within their quoted uncertainties induce relative variations in $\dot{\omega}$ for the two targets dominating our fits, V541\,Cyg and AL\,Dor, of lower than 1\%.

With new {\em TESS} data still to come, observing new systems and increasing the data coverage of those already observed, promising systems still without a detection of apsidal motion could be added to further extend our study. In order to perform even more stringent tests of gravitation theories, an increase in the number of systems with a GR apsidal motion relative contribution larger than the threshold used in this work is needed. We therefore encourage spectroscopic monitoring and precise determination of absolute properties of long-period, eccentric eclipsing binaries with {\em TESS} photometric data. 

\begin{acknowledgements}
Ra\"ul Vera (EHU, Spain) is gratefully acknowledged for his guidance and discussions regarding the PPN formalism and observational parameter constraints. Antonio Claret (IAA, Spain) is gratefully acknowledged for providing internal structure constants specifically calculated for our eclipsing binary systems. This paper includes data collected by the {\em TESS} mission. Funding for the {\em TESS} mission is provided by the NASA Explorer Program. We acknowledge support from the Spanish Ministry of Science and Innovation and the European Regional Development Fund through grant PGC2018-098153-B-C33, the support of the Generalitat de Catalunya/CERCA programme, and the Ag\`encia de Gesti\'o d'Ajuts Universitaris i de Recerca of the Generalitat de Catalunya, with additional funding from the European FEDER/ERF funds, \emph{L'FSE inverteix en el teu futur}. This work has been carried out within the framework of the PhD programme in Physics of the Universitat Aut\`onoma de Barcelona.
\end{acknowledgements}

\bibliographystyle{aa} 
\bibliography{bibtex.bib}

\begin{thebibliography}{101}
\expandafter\ifx\csname natexlab\endcsname\relax\def\natexlab#1{#1}\fi

\bibitem[{{Agerer} {et~al.}(1994){Agerer}, {Busch}, {Kleikamp}, \&
  {Moschner}}]{Agerer1994}
{Agerer}, F., {Busch}, H., {Kleikamp}, W., \& {Moschner}, W. 1994, Berliner
  Arbeitsgemeinschaft fuer Veraenderliche Sterne - Mitteilungen, 69, 1

\bibitem[{{Aigrain} {et~al.}(2016){Aigrain}, {Parviainen}, \&
  {Pope}}]{Aigrain2016}
{Aigrain}, S., {Parviainen}, H., \& {Pope}, B.~J.~S. 2016, \mnras, 459, 2408

\bibitem[{{Albrecht} {et~al.}(2009){Albrecht}, {Reffert}, {Snellen}, \&
  {Winn}}]{Albrecht2009}
{Albrecht}, S., {Reffert}, S., {Snellen}, I. A.~G., \& {Winn}, J.~N. 2009,
  \nat, 461, 373

\bibitem[{{Albrecht} {et~al.}(2013){Albrecht}, {Setiawan}, {Torres},
  {Fabrycky}, \& {Winn}}]{Albrecht2013}
{Albrecht}, S., {Setiawan}, J., {Torres}, G., {Fabrycky}, D.~C., \& {Winn},
  J.~N. 2013, \apj, 767, 32

\bibitem[{{Andersen}(1991)}]{Andersen1991}
{Andersen}, J. 1991, \aapr, 3, 91

\bibitem[{{Andersen} {et~al.}(1987){Andersen}, {Clausen}, \&
  {Nordstrom}}]{Andersen1987}
{Andersen}, J., {Clausen}, J.~V., \& {Nordstrom}, B. 1987, \aap, 175, 60

\bibitem[{{Andersen} \& {Vaz}(1984)}]{Andersen1984}
{Andersen}, J., \& {Vaz}, L.~P.~R. 1984, \aap, 130, 102

\bibitem[{{Baker} {et~al.}(2015){Baker}, {Psaltis}, \& {Skordis}}]{Baker2015}
{Baker}, T., {Psaltis}, D., \& {Skordis}, C. 2015, \apj, 802, 63

\bibitem[{{Bak{\i}{\textcommabelow s}} {et~al.}(2008){Bak{\i}{\textcommabelow
  s}}, {Bak{\i}{\textcommabelow s}}, {Demircan}, \& {Eker}}]{Bakis2008}
{Bak{\i}{\textcommabelow s}}, V., {Bak{\i}{\textcommabelow s}}, H., {Demircan},
  O., \& {Eker}, Z. 2008, \mnras, 384, 1657

\bibitem[{{Bertotti} {et~al.}(2003){Bertotti}, {Iess}, \&
  {Tortora}}]{Bertotti2003}
{Bertotti}, B., {Iess}, L., \& {Tortora}, P. 2003, \nat, 425, 374

\bibitem[{{Breinhorst} {et~al.}(1973){Breinhorst}, {Pfleiderer}, {Reinhardt},
  \& {Karimie}}]{Breinhorst1973}
{Breinhorst}, R.~A., {Pfleiderer}, J., {Reinhardt}, M., \& {Karimie}, M.~T.
  1973, \aap, 22, 239

\bibitem[{{Claret}(1997)}]{Claret1997}
{Claret}, A. 1997, \aap, 327, 11

\bibitem[{{Claret} \& {Gim{\'e}nez}(1993)}]{Claret1993}
{Claret}, A., \& {Gim{\'e}nez}, A. 1993, \aap, 277, 487

\bibitem[{{Claret} \& {Gim{\'e}nez}(2010)}]{Claret2010b}
{Claret}, A., \& {Gim{\'e}nez}, A. 2010, \aap, 519, A57

\bibitem[{{Claret} \& {Torres}(2017)}]{Claret2017b}
{Claret}, A., \& {Torres}, G. 2017, \apj, 849, 18

\bibitem[{{Claret} \& {Torres}(2018)}]{Claret2018b}
{Claret}, A., \& {Torres}, G. 2018, \apj, 859, 100

\bibitem[{{Claret} \& {Torres}(2019)}]{Claret2019b}
{Claret}, A., \& {Torres}, G. 2019, \apj, 876, 134

\bibitem[{{Claret} {et~al.}(2010){Claret}, {Torres}, \& {Wolf}}]{Claret2010}
{Claret}, A., {Torres}, G., \& {Wolf}, M. 2010, \aap, 515, A4

\bibitem[{{Dariush} {et~al.}(2006){Dariush}, {Mosleh}, \&
  {Dariush}}]{Dariush2006}
{Dariush}, A., {Mosleh}, M., \& {Dariush}, D. 2006, \apss, 305, 85

\bibitem[{{Davies}(2007)}]{Davies2007}
{Davies}, D. 2007, Peremennye Zvezdy Prilozhenie, 7, 16

\bibitem[{{De Laurentis} {et~al.}(2012){De Laurentis}, {De Rosa}, {Garufi}, \&
  {Milano}}]{DeLaurentis2012}
{De Laurentis}, M., {De Rosa}, R., {Garufi}, F., \& {Milano}, L. 2012, \mnras,
  424, 2371

\bibitem[{{Diethelm}(1992)}]{Diethelm1992}
{Diethelm}, R. 1992, Bulletin der Bedeckungsveraenderlichen-Beobachter der
  Schweizerischen Astronomischen Gesellschaft, 99, 10

\bibitem[{{Estabrook}(1969)}]{Estabrook1969}
{Estabrook}, F.~B. 1969, \apj, 158, 81

\bibitem[{{Feiden} \& {Chaboyer}(2012)}]{Feiden2012}
{Feiden}, G.~A., \& {Chaboyer}, B. 2012, \apj, 757, 42

\bibitem[{{Feinstein} {et~al.}(2019){Feinstein}, {Montet}, {Foreman-Mackey},
  {Bedell}, {Saunders}, {Bean}, {Christiansen}, {Hedges}, {Luger}, {Scolnic},
  \& {Cardoso}}]{Feinstein2019}
{Feinstein}, A.~D., {Montet}, B.~T., {Foreman-Mackey}, D., {et~al.} 2019,
  \pasp, 131, 094502

\bibitem[{{Ferrero} {et~al.}(2013){Ferrero}, {Gamen}, {Benvenuto}, \&
  {Fern{\'a}ndez-Laj{\'u}s}}]{Ferrero2013}
{Ferrero}, G., {Gamen}, R., {Benvenuto}, O., \& {Fern{\'a}ndez-Laj{\'u}s}, E.
  2013, \mnras, 433, 1300

\bibitem[{{Foreman-Mackey}(2015)}]{Foreman2015}
{Foreman-Mackey}, D. 2015, {George: Gaussian Process regression}

\bibitem[{{Gaia Collaboration} {et~al.}(2018){Gaia Collaboration}, {Brown},
  {Vallenari}, {Prusti}, {de Bruijne}, {Babusiaux}, {Bailer-Jones}, {Biermann},
  {Evans}, {Eyer}, {Jansen}, {Jordi}, {Klioner}, {Lammers}, {Lindegren},
  {Luri}, {Mignard}, {Panem}, {Pourbaix}, {Randich}, {Sartoretti}, {Siddiqui},
  {Soubiran}, {van Leeuwen}, {Walton}, {Arenou}, {Bastian}, {Cropper},
  {Drimmel}, {Katz}, {Lattanzi}, {Bakker}, {Cacciari}, {Casta{\~n}eda},
  {Chaoul}, {Cheek}, {De Angeli}, {Fabricius}, {Guerra}, {Holl}, {Masana},
  {Messineo}, {Mowlavi}, {Nienartowicz}, {Panuzzo}, {Portell}, {Riello},
  {Seabroke}, {Tanga}, {Th{\'e}venin}, {Gracia-Abril}, {Comoretto},
  {Garcia-Reinaldos}, {Teyssier}, {Altmann}, {Andrae}, {Audard},
  {Bellas-Velidis}, {Benson}, {Berthier}, {Blomme}, {Burgess}, {Busso},
  {Carry}, {Cellino}, {Clementini}, {Clotet}, {Creevey}, {Davidson}, {De
  Ridder}, {Delchambre}, {Dell'Oro}, {Ducourant},
  {Fern{\'a}ndez-Hern{\'a}ndez}, {Fouesneau}, {Fr{\'e}mat}, {Galluccio},
  {Garc{\'\i}a-Torres}, {Gonz{\'a}lez-N{\'u}{\~n}ez}, {Gonz{\'a}lez-Vidal},
  {Gosset}, {Guy}, {Halbwachs}, {Hambly}, {Harrison}, {Hern{\'a}ndez},
  {Hestroffer}, {Hodgkin}, {Hutton}, {Jasniewicz}, {Jean-Antoine-Piccolo},
  {Jordan}, {Korn}, {Krone-Martins}, {Lanzafame}, {Lebzelter}, {L{\"o}ffler},
  {Manteiga}, {Marrese}, {Mart{\'\i}n-Fleitas}, {Moitinho}, {Mora}, {Muinonen},
  {Osinde}, {Pancino}, {Pauwels}, {Petit}, {Recio-Blanco}, {Richards},
  {Rimoldini}, {Robin}, {Sarro}, {Siopis}, {Smith}, {Sozzetti}, {S{\"u}veges},
  {Torra}, {van Reeven}, {Abbas}, {Abreu Aramburu}, {Accart}, {Aerts},
  {Altavilla}, {{\'A}lvarez}, {Alvarez}, {Alves}, {Anderson}, {Andrei},
  {Anglada Varela}, {Antiche}, {Antoja}, {Arcay}, {Astraatmadja}, {Bach},
  {Baker}, {Balaguer-N{\'u}{\~n}ez}, {Balm}, {Barache}, {Barata}, {Barbato},
  {Barblan}, {Barklem}, {Barrado}, {Barros}, {Barstow}, {Bartholom{\'e}
  Mu{\~n}oz}, {Bassilana}, {Becciani}, {Bellazzini}, {Berihuete}, {Bertone},
  {Bianchi}, {Bienaym{\'e}}, {Blanco-Cuaresma}, {Boch}, {Boeche}, {Bombrun},
  {Borrachero}, {Bossini}, {Bouquillon}, {Bourda}, {Bragaglia}, {Bramante},
  {Breddels}, {Bressan}, {Brouillet}, {Br{\"u}semeister}, {Brugaletta},
  {Bucciarelli}, {Burlacu}, {Busonero}, {Butkevich}, {Buzzi}, {Caffau},
  {Cancelliere}, {Cannizzaro}, {Cantat-Gaudin}, {Carballo}, {Carlucci},
  {Carrasco}, {Casamiquela}, {Castellani}, {Castro-Ginard}, {Charlot},
  {Chemin}, {Chiavassa}, {Cocozza}, {Costigan}, {Cowell}, {Crifo}, {Crosta},
  {Crowley}, {Cuypers}, {Dafonte}, {Damerdji}, {Dapergolas}, {David}, {David},
  {de Laverny}, {De Luise}, {De March}, {de Martino}, {de Souza}, {de Torres},
  {Debosscher}, {del Pozo}, {Delbo}, {Delgado}, {Delgado}, {Di Matteo},
  {Diakite}, {Diener}, {Distefano}, {Dolding}, {Drazinos}, {Dur{\'a}n},
  {Edvardsson}, {Enke}, {Eriksson}, {Esquej}, {Eynard Bontemps}, {Fabre},
  {Fabrizio}, {Faigler}, {Falc{\~a}o}, {Farr{\`a}s Casas}, {Federici},
  {Fedorets}, {Fernique}, {Figueras}, {Filippi}, {Findeisen}, {Fonti},
  {Fraile}, {Fraser}, {Fr{\'e}zouls}, {Gai}, {Galleti}, {Garabato},
  {Garc{\'\i}a-Sedano}, {Garofalo}, {Garralda}, {Gavel}, {Gavras}, {Gerssen},
  {Geyer}, {Giacobbe}, {Gilmore}, {Girona}, {Giuffrida}, {Glass}, {Gomes},
  {Granvik}, {Gueguen}, {Guerrier}, {Guiraud}, {Guti{\'e}rrez-S{\'a}nchez},
  {Haigron}, {Hatzidimitriou}, {Hauser}, {Haywood}, {Heiter}, {Helmi}, {Heu},
  {Hilger}, {Hobbs}, {Hofmann}, {Holland}, {Huckle}, {Hypki}, {Icardi},
  {Jan{\ss}en}, {Jevardat de Fombelle}, {Jonker}, {Juh{\'a}sz}, {Julbe},
  {Karampelas}, {Kewley}, {Klar}, {Kochoska}, {Kohley}, {Kolenberg},
  {Kontizas}, {Kontizas}, {Koposov}, {Kordopatis}, {Kostrzewa-Rutkowska},
  {Koubsky}, {Lambert}, {Lanza}, {Lasne}, {Lavigne}, {Le Fustec}, {Le
  Poncin-Lafitte}, {Lebreton}, {Leccia}, {Leclerc}, {Lecoeur-Taibi},
  {Lenhardt}, {Leroux}, {Liao}, {Licata}, {Lindstr{\o}m}, {Lister}, {Livanou},
  {Lobel}, {L{\'o}pez}, {Managau}, {Mann}, {Mantelet}, {Marchal}, {Marchant},
  {Marconi}, {Marinoni}, {Marschalk{\'o}}, {Marshall}, {Martino}, {Marton},
  {Mary}, {Massari}, {Matijevi{\v{c}}}, {Mazeh}, {McMillan}, {Messina},
  {Michalik}, {Millar}, {Molina}, {Molinaro}, {Moln{\'a}r}, {Montegriffo},
  {Mor}, {Morbidelli}, {Morel}, {Morris}, {Mulone}, {Muraveva}, {Musella},
  {Nelemans}, {Nicastro}, {Noval}, {O'Mullane}, {Ord{\'e}novic},
  {Ord{\'o}{\~n}ez-Blanco}, {Osborne}, {Pagani}, {Pagano}, {Pailler},
  {Palacin}, {Palaversa}, {Panahi}, {Pawlak}, {Piersimoni}, {Pineau}, {Plachy},
  {Plum}, {Poggio}, {Poujoulet}, {Pr{\v{s}}a}, {Pulone}, {Racero}, {Ragaini},
  {Rambaux}, {Ramos-Lerate}, {Regibo}, {Reyl{\'e}}, {Riclet}, {Ripepi}, {Riva},
  {Rivard}, {Rixon}, {Roegiers}, {Roelens}, {Romero-G{\'o}mez}, {Rowell},
  {Royer}, {Ruiz-Dern}, {Sadowski}, {Sagrist{\`a} Sell{\'e}s}, {Sahlmann},
  {Salgado}, {Salguero}, {Sanna}, {Santana-Ros}, {Sarasso}, {Savietto},
  {Schultheis}, {Sciacca}, {Segol}, {Segovia}, {S{\'e}gransan}, {Shih},
  {Siltala}, {Silva}, {Smart}, {Smith}, {Solano}, {Solitro}, {Sordo}, {Soria
  Nieto}, {Souchay}, {Spagna}, {Spoto}, {Stampa}, {Steele},
  {Steidelm{\"u}ller}, {Stephenson}, {Stoev}, {Suess}, {Surdej}, {Szabados},
  {Szegedi-Elek}, {Tapiador}, {Taris}, {Tauran}, {Taylor}, {Teixeira},
  {Terrett}, {Teyssandier}, {Thuillot}, {Titarenko}, {Torra Clotet}, {Turon},
  {Ulla}, {Utrilla}, {Uzzi}, {Vaillant}, {Valentini}, {Valette}, {van Elteren},
  {Van Hemelryck}, {van Leeuwen}, {Vaschetto}, {Vecchiato}, {Veljanoski},
  {Viala}, {Vicente}, {Vogt}, {von Essen}, {Voss}, {Votruba}, {Voutsinas},
  {Walmsley}, {Weiler}, {Wertz}, {Wevers}, {Wyrzykowski}, {Yoldas},
  {{\v{Z}}erjal}, {Ziaeepour}, {Zorec}, {Zschocke}, {Zucker}, {Zurbach}, \&
  {Zwitter}}]{Gaia2018}
{Gaia Collaboration}, {Brown}, A.~G.~A., {Vallenari}, A., {et~al.} 2018, \aap,
  616, A1

\bibitem[{{Gaia Collaboration} {et~al.}(2016){Gaia Collaboration}, {Prusti},
  {de Bruijne}, {Brown}, {Vallenari}, {Babusiaux}, {Bailer-Jones}, {Bastian},
  {Biermann}, {Evans}, {Eyer}, {Jansen}, {Jordi}, {Klioner}, {Lammers},
  {Lindegren}, {Luri}, {Mignard}, {Milligan}, {Panem}, {Poinsignon},
  {Pourbaix}, {Randich}, {Sarri}, {Sartoretti}, {Siddiqui}, {Soubiran},
  {Valette}, {van Leeuwen}, {Walton}, {Aerts}, {Arenou}, {Cropper}, {Drimmel},
  {H{\o}g}, {Katz}, {Lattanzi}, {O'Mullane}, {Grebel}, {Holland}, {Huc},
  {Passot}, {Bramante}, {Cacciari}, {Casta{\~n}eda}, {Chaoul}, {Cheek}, {De
  Angeli}, {Fabricius}, {Guerra}, {Hern{\'a}ndez}, {Jean-Antoine-Piccolo},
  {Masana}, {Messineo}, {Mowlavi}, {Nienartowicz}, {Ord{\'o}{\~n}ez-Blanco},
  {Panuzzo}, {Portell}, {Richards}, {Riello}, {Seabroke}, {Tanga},
  {Th{\'e}venin}, {Torra}, {Els}, {Gracia-Abril}, {Comoretto},
  {Garcia-Reinaldos}, {Lock}, {Mercier}, {Altmann}, {Andrae}, {Astraatmadja},
  {Bellas-Velidis}, {Benson}, {Berthier}, {Blomme}, {Busso}, {Carry},
  {Cellino}, {Clementini}, {Cowell}, {Creevey}, {Cuypers}, {Davidson}, {De
  Ridder}, {de Torres}, {Delchambre}, {Dell'Oro}, {Ducourant}, {Fr{\'e}mat},
  {Garc{\'\i}a-Torres}, {Gosset}, {Halbwachs}, {Hambly}, {Harrison}, {Hauser},
  {Hestroffer}, {Hodgkin}, {Huckle}, {Hutton}, {Jasniewicz}, {Jordan},
  {Kontizas}, {Korn}, {Lanzafame}, {Manteiga}, {Moitinho}, {Muinonen},
  {Osinde}, {Pancino}, {Pauwels}, {Petit}, {Recio-Blanco}, {Robin}, {Sarro},
  {Siopis}, {Smith}, {Smith}, {Sozzetti}, {Thuillot}, {van Reeven}, {Viala},
  {Abbas}, {Abreu Aramburu}, {Accart}, {Aguado}, {Allan}, {Allasia},
  {Altavilla}, {{\'A}lvarez}, {Alves}, {Anderson}, {Andrei}, {Anglada Varela},
  {Antiche}, {Antoja}, {Ant{\'o}n}, {Arcay}, {Atzei}, {Ayache}, {Bach},
  {Baker}, {Balaguer-N{\'u}{\~n}ez}, {Barache}, {Barata}, {Barbier}, {Barblan},
  {Baroni}, {Barrado y Navascu{\'e}s}, {Barros}, {Barstow}, {Becciani},
  {Bellazzini}, {Bellei}, {Bello Garc{\'\i}a}, {Belokurov}, {Bendjoya},
  {Berihuete}, {Bianchi}, {Bienaym{\'e}}, {Billebaud}, {Blagorodnova},
  {Blanco-Cuaresma}, {Boch}, {Bombrun}, {Borrachero}, {Bouquillon}, {Bourda},
  {Bouy}, {Bragaglia}, {Breddels}, {Brouillet}, {Br{\"u}semeister},
  {Bucciarelli}, {Budnik}, {Burgess}, {Burgon}, {Burlacu}, {Busonero}, {Buzzi},
  {Caffau}, {Cambras}, {Campbell}, {Cancelliere}, {Cantat-Gaudin}, {Carlucci},
  {Carrasco}, {Castellani}, {Charlot}, {Charnas}, {Charvet}, {Chassat},
  {Chiavassa}, {Clotet}, {Cocozza}, {Collins}, {Collins}, {Costigan}, {Crifo},
  {Cross}, {Crosta}, {Crowley}, {Dafonte}, {Damerdji}, {Dapergolas}, {David},
  {David}, {De Cat}, {de Felice}, {de Laverny}, {De Luise}, {De March}, {de
  Martino}, {de Souza}, {Debosscher}, {del Pozo}, {Delbo}, {Delgado},
  {Delgado}, {di Marco}, {Di Matteo}, {Diakite}, {Distefano}, {Dolding}, {Dos
  Anjos}, {Drazinos}, {Dur{\'a}n}, {Dzigan}, {Ecale}, {Edvardsson}, {Enke},
  {Erdmann}, {Escolar}, {Espina}, {Evans}, {Eynard Bontemps}, {Fabre},
  {Fabrizio}, {Faigler}, {Falc{\~a}o}, {Farr{\`a}s Casas}, {Faye}, {Federici},
  {Fedorets}, {Fern{\'a}ndez-Hern{\'a}ndez}, {Fernique}, {Fienga}, {Figueras},
  {Filippi}, {Findeisen}, {Fonti}, {Fouesneau}, {Fraile}, {Fraser}, {Fuchs},
  {Furnell}, {Gai}, {Galleti}, {Galluccio}, {Garabato}, {Garc{\'\i}a-Sedano},
  {Gar{\'e}}, {Garofalo}, {Garralda}, {Gavras}, {Gerssen}, {Geyer}, {Gilmore},
  {Girona}, {Giuffrida}, {Gomes}, {Gonz{\'a}lez-Marcos},
  {Gonz{\'a}lez-N{\'u}{\~n}ez}, {Gonz{\'a}lez-Vidal}, {Granvik}, {Guerrier},
  {Guillout}, {Guiraud}, {G{\'u}rpide}, {Guti{\'e}rrez-S{\'a}nchez}, {Guy},
  {Haigron}, {Hatzidimitriou}, {Haywood}, {Heiter}, {Helmi}, {Hobbs},
  {Hofmann}, {Holl}, {Holland}, {Hunt}, {Hypki}, {Icardi}, {Irwin}, {Jevardat
  de Fombelle}, {Jofr{\'e}}, {Jonker}, {Jorissen}, {Julbe}, {Karampelas},
  {Kochoska}, {Kohley}, {Kolenberg}, {Kontizas}, {Koposov}, {Kordopatis},
  {Koubsky}, {Kowalczyk}, {Krone-Martins}, {Kudryashova}, {Kull}, {Bachchan},
  {Lacoste-Seris}, {Lanza}, {Lavigne}, {Le Poncin-Lafitte}, {Lebreton},
  {Lebzelter}, {Leccia}, {Leclerc}, {Lecoeur-Taibi}, {Lemaitre}, {Lenhardt},
  {Leroux}, {Liao}, {Licata}, {Lindstr{\o}m}, {Lister}, {Livanou}, {Lobel},
  {L{\"o}ffler}, {L{\'o}pez}, {Lopez-Lozano}, {Lorenz}, {Loureiro},
  {MacDonald}, {Magalh{\~a}es Fernandes}, {Managau}, {Mann}, {Mantelet},
  {Marchal}, {Marchant}, {Marconi}, {Marie}, {Marinoni}, {Marrese},
  {Marschalk{\'o}}, {Marshall}, {Mart{\'\i}n-Fleitas}, {Martino}, {Mary},
  {Matijevi{\v{c}}}, {Mazeh}, {McMillan}, {Messina}, {Mestre}, {Michalik},
  {Millar}, {Miranda}, {Molina}, {Molinaro}, {Molinaro}, {Moln{\'a}r},
  {Moniez}, {Montegriffo}, {Monteiro}, {Mor}, {Mora}, {Morbidelli}, {Morel},
  {Morgenthaler}, {Morley}, {Morris}, {Mulone}, {Muraveva}, {Musella},
  {Narbonne}, {Nelemans}, {Nicastro}, {Noval}, {Ord{\'e}novic},
  {Ordieres-Mer{\'e}}, {Osborne}, {Pagani}, {Pagano}, {Pailler}, {Palacin},
  {Palaversa}, {Parsons}, {Paulsen}, {Pecoraro}, {Pedrosa}, {Pentik{\"a}inen},
  {Pereira}, {Pichon}, {Piersimoni}, {Pineau}, {Plachy}, {Plum}, {Poujoulet},
  {Pr{\v{s}}a}, {Pulone}, {Ragaini}, {Rago}, {Rambaux}, {Ramos-Lerate},
  {Ranalli}, {Rauw}, {Read}, {Regibo}, {Renk}, {Reyl{\'e}}, {Ribeiro},
  {Rimoldini}, {Ripepi}, {Riva}, {Rixon}, {Roelens}, {Romero-G{\'o}mez},
  {Rowell}, {Royer}, {Rudolph}, {Ruiz-Dern}, {Sadowski}, {Sagrist{\`a}
  Sell{\'e}s}, {Sahlmann}, {Salgado}, {Salguero}, {Sarasso}, {Savietto},
  {Schnorhk}, {Schultheis}, {Sciacca}, {Segol}, {Segovia}, {Segransan},
  {Serpell}, {Shih}, {Smareglia}, {Smart}, {Smith}, {Solano}, {Solitro},
  {Sordo}, {Soria Nieto}, {Souchay}, {Spagna}, {Spoto}, {Stampa}, {Steele},
  {Steidelm{\"u}ller}, {Stephenson}, {Stoev}, {Suess}, {S{\"u}veges}, {Surdej},
  {Szabados}, {Szegedi-Elek}, {Tapiador}, {Taris}, {Tauran}, {Taylor},
  {Teixeira}, {Terrett}, {Tingley}, {Trager}, {Turon}, {Ulla}, {Utrilla},
  {Valentini}, {van Elteren}, {Van Hemelryck}, {van Leeuwen}, {Varadi},
  {Vecchiato}, {Veljanoski}, {Via}, {Vicente}, {Vogt}, {Voss}, {Votruba},
  {Voutsinas}, {Walmsley}, {Weiler}, {Weingrill}, {Werner}, {Wevers},
  {Whitehead}, {Wyrzykowski}, {Yoldas}, {{\v{Z}}erjal}, {Zucker}, {Zurbach},
  {Zwitter}, {Alecu}, {Allen}, {Allende Prieto}, {Amorim},
  {Anglada-Escud{\'e}}, {Arsenijevic}, {Azaz}, {Balm}, {Beck}, {Bernstein},
  {Bigot}, {Bijaoui}, {Blasco}, {Bonfigli}, {Bono}, {Boudreault}, {Bressan},
  {Brown}, {Brunet}, {Bunclark}, {Buonanno}, {Butkevich}, {Carret}, {Carrion},
  {Chemin}, {Ch{\'e}reau}, {Corcione}, {Darmigny}, {de Boer}, {de Teodoro}, {de
  Zeeuw}, {Delle Luche}, {Domingues}, {Dubath}, {Fodor}, {Fr{\'e}zouls},
  {Fries}, {Fustes}, {Fyfe}, {Gallardo}, {Gallegos}, {Gardiol}, {Gebran},
  {Gomboc}, {G{\'o}mez}, {Grux}, {Gueguen}, {Heyrovsky}, {Hoar}, {Iannicola},
  {Isasi Parache}, {Janotto}, {Joliet}, {Jonckheere}, {Keil}, {Kim},
  {Klagyivik}, {Klar}, {Knude}, {Kochukhov}, {Kolka}, {Kos}, {Kutka}, {Lainey},
  {LeBouquin}, {Liu}, {Loreggia}, {Makarov}, {Marseille}, {Martayan},
  {Martinez-Rubi}, {Massart}, {Meynadier}, {Mignot}, {Munari}, {Nguyen},
  {Nordlander}, {Ocvirk}, {O'Flaherty}, {Olias Sanz}, {Ortiz}, {Osorio},
  {Oszkiewicz}, {Ouzounis}, {Palmer}, {Park}, {Pasquato}, {Peltzer}, {Peralta},
  {P{\'e}turaud}, {Pieniluoma}, {Pigozzi}, {Poels}, {Prat}, {Prod'homme},
  {Raison}, {Rebordao}, {Risquez}, {Rocca-Volmerange}, {Rosen}, {Ruiz-Fuertes},
  {Russo}, {Sembay}, {Serraller Vizcaino}, {Short}, {Siebert}, {Silva},
  {Sinachopoulos}, {Slezak}, {Soffel}, {Sosnowska}, {Strai{\v{z}}ys}, {ter
  Linden}, {Terrell}, {Theil}, {Tiede}, {Troisi}, {Tsalmantza}, {Tur},
  {Vaccari}, {Vachier}, {Valles}, {Van Hamme}, {Veltz}, {Virtanen}, {Wallut},
  {Wichmann}, {Wilkinson}, {Ziaeepour}, \& {Zschocke}}]{Gaia2016}
{Gaia Collaboration}, {Prusti}, T., {de Bruijne}, J.~H.~J., {et~al.} 2016,
  \aap, 595, A1

\bibitem[{{Gallenne} {et~al.}(2019){Gallenne}, {Pietrzy{\'n}ski}, {Graczyk},
  {Pilecki}, {Storm}, {Nardetto}, {Taormina}, {Gieren}, {Tkachenko},
  {Kervella}, {M{\'e}rand }, \& {Weber}}]{Gallenne2019}
{Gallenne}, A., {Pietrzy{\'n}ski}, G., {Graczyk}, D., {et~al.} 2019, \aap, 632,
  A31

\bibitem[{{Gibson} {et~al.}(2011){Gibson}, {Pont}, \& {Aigrain}}]{Gibson2011}
{Gibson}, N.~P., {Pont}, F., \& {Aigrain}, S. 2011, \mnras, 411, 2199

\bibitem[{{Gim{\'e}nez}(1985)}]{Gimenez1985}
{Gim{\'e}nez}, A. 1985, \apj, 297, 405

\bibitem[{{Gim{\'e}nez}(2007)}]{Gimenez2007}
{Gim{\'e}nez}, A. 2007, in IAU Symposium, Vol. 240, Binary Stars as Critical
  Tools \& Tests in Contemporary Astrophysics, ed. W.~I. {Hartkopf},
  P.~{Harmanec}, \& E.~F. {Guinan}, 290--298

\bibitem[{{Gim{\'e}nez} \& {Bastero}(1995)}]{Gimenez1995}
{Gim{\'e}nez}, A., \& {Bastero}, M. 1995, \apss, 226, 99

\bibitem[{{Gim{\'e}nez} {et~al.}(1987){Gim{\'e}nez}, {Kim}, \&
  {Nha}}]{Gimenez1987}
{Gim{\'e}nez}, A., {Kim}, C.-H., \& {Nha}, I.-S. 1987, \mnras, 224, 543

\bibitem[{{Graczyk} {et~al.}(2019){Graczyk}, {Pietrzy{\'n}ski}, {Gieren},
  {Storm}, {Nardetto}, {Gallenne}, {Maxted}, {Kervella}, {Ko{\l}aczkowski},
  {Konorski}, {Pilecki}, {Zgirski}, {G{\'o}rski}, {Suchomska}, {Karczmarek},
  {Taormina}, {Wielg{\'o}rski}, {Narloch}, {Smolec}, {Chini}, \&
  {Breuval}}]{Graczyk2019}
{Graczyk}, D., {Pietrzy{\'n}ski}, G., {Gieren}, W., {et~al.} 2019, \apj, 872,
  85

\bibitem[{{Guinan} {et~al.}(1996){Guinan}, {Maley}, \& {Marshall}}]{Guinan1996}
{Guinan}, E.~F., {Maley}, J.~A., \& {Marshall}, J.~J. 1996, IBVS, 4362, 1

\bibitem[{{Guinan} \& {Maloney}(1985)}]{Guinan1985}
{Guinan}, E.~F., \& {Maloney}, F.~P. 1985, \aj, 90, 1519

\bibitem[{{Harmanec} {et~al.}(2014){Harmanec}, {Holmgren}, {Wolf},
  {Bo{\v{z}}i{\'c}}, {Guinan}, {Kang}, {Mayer}, {McCook}, {Nemravov{\'a}},
  {Yang}, {{\v{S}}lechta}, {Ru{\v{z}}djak}, {Sudar}, \&
  {Svoboda}}]{Harmanec2014}
{Harmanec}, P., {Holmgren}, D.~E., {Wolf}, M., {et~al.} 2014, \aap, 563, A120

\bibitem[{{Hejlesen}(1987)}]{Hejlesen1987}
{Hejlesen}, P.~M. 1987, \aaps, 69, 251

\bibitem[{{Higl} \& {Weiss}(2017)}]{Higl2017}
{Higl}, J., \& {Weiss}, A. 2017, \aap, 608, A62

\bibitem[{{Hubscher}(2015)}]{Hubscher2015a}
{Hubscher}, J. 2015, IBVS, 6152, 1

\bibitem[{{Hubscher} \& {Lehmann}(2015)}]{Hubscher2015b}
{Hubscher}, J., \& {Lehmann}, P.~B. 2015, IBVS, 6149, 1

\bibitem[{{Hubscher} {et~al.}(2005){Hubscher}, {Paschke}, \&
  {Walter}}]{Hubscher2005}
{Hubscher}, J., {Paschke}, A., \& {Walter}, F. 2005, Information Bulletin on
  Variable Stars, 5657, 1

\bibitem[{{Hut}(1981)}]{Hut1981}
{Hut}, P. 1981, \aap, 99, 126

\bibitem[{{Jenkins} {et~al.}(2016){Jenkins}, {Twicken}, {McCauliff},
  {Campbell}, {Sanderfer}, {Lung}, {Mansouri-Samani}, {Girouard}, {Tenenbaum},
  {Klaus}, {Smith}, {Caldwell}, {Chacon}, {Henze}, {Heiges}, {Latham},
  {Morgan}, {Swade}, {Rinehart}, \& {Vanderspek}}]{Jenkins2016}
{Jenkins}, J.~M., {Twicken}, J.~D., {McCauliff}, S., {et~al.} 2016, in Society
  of Photo-Optical Instrumentation Engineers (SPIE) Conference Series, Vol.
  9913, Software and Cyberinfrastructure for Astronomy IV, 99133E

\bibitem[{{Karpowicz}(1961)}]{Karpowicz1961}
{Karpowicz}, M. 1961, \actaa, 11, 51

\bibitem[{{Khaliullin}(1985)}]{Khaliullin1985}
{Khaliullin}, K.~F. 1985, \apj, 299, 668

\bibitem[{{Khaliullin} \& {Khaliullina}(2007)}]{Khaliullin2007}
{Khaliullin}, K.~F., \& {Khaliullina}, A.~I. 2007, \mnras, 382, 356

\bibitem[{{Kim} {et~al.}(2018){Kim}, {Kreiner}, {Zakrzewski}, {Og{\l}oza},
  {Kim}, \& {Jeong}}]{Kim2018}
{Kim}, C.~H., {Kreiner}, J.~M., {Zakrzewski}, B., {et~al.} 2018, \apjs, 235, 41

\bibitem[{{Kopal}(1959)}]{Kopal1959}
{Kopal}, Z. 1959, {Close binary systems} (New York: John Wiley \& Sons Inc.)

\bibitem[{{Kopal}(1978)}]{Kopal1978}
{Kopal}, Z. 1978, {Dynamics of close binary systems} (Springer Netherlands)

\bibitem[{{Kozai}(1962)}]{Kozai1962}
{Kozai}, Y. 1962, \aj, 67, 591

\bibitem[{{Kwee} \& {van Woerden}(1956)}]{Kwee1956}
{Kwee}, K.~K., \& {van Woerden}, H. 1956, \bain, 12, 327

\bibitem[{{Lacy}(1998)}]{Lacy1998}
{Lacy}, C. H.~S. 1998, \aj, 115, 801

\bibitem[{{Lacy}(2002)}]{Lacy2002IBVS}
{Lacy}, C.~H.~S. 2002, Information Bulletin on Variable Stars, 5357, 1

\bibitem[{{Lacy}(2004)}]{Lacy2004IBVS}
{Lacy}, C.~H.~S. 2004, Information Bulletin on Variable Stars, 5577, 1

\bibitem[{{Lacy} {et~al.}(2004){Lacy}, {Claret}, \& {Sabby}}]{Lacy2004}
{Lacy}, C. H.~S., {Claret}, A., \& {Sabby}, J.~A. 2004, \aj, 128, 1340

\bibitem[{{Lacy} {et~al.}(2001){Lacy}, {Hood}, \& {Straughn}}]{Lacy2001IBVS}
{Lacy}, C.~H.~S., {Hood}, B., \& {Straughn}, A. 2001, Information Bulletin on
  Variable Stars, 5067, 1

\bibitem[{{Lacy} {et~al.}(2006){Lacy}, {Torres}, {Claret}, \&
  {Menke}}]{Lacy2006}
{Lacy}, C. H.~S., {Torres}, G., {Claret}, A., \& {Menke}, J.~L. 2006, \aj, 131,
  2664

\bibitem[{{Lacy} {et~al.}(2005){Lacy}, {Torres}, {Claret}, \& {Vaz}}]{Lacy2005}
{Lacy}, C. H.~S., {Torres}, G., {Claret}, A., \& {Vaz}, L. P.~R. 2005, \aj,
  130, 2838

\bibitem[{{Lastennet} \& {Valls-Gabaud}(2002)}]{Lastennet2002}
{Lastennet}, E., \& {Valls-Gabaud}, D. 2002, \aap, 396, 551

\bibitem[{Levi-Civita(1937)}]{Levi1937}
Levi-Civita, T. 1937, Am. J. Math., 59, 225

\bibitem[{{Lidov}(1962)}]{Lidov1962}
{Lidov}, M.~L. 1962, \planss, 9, 719

\bibitem[{{Mikul{\'a}{\v{s}}ek} {et~al.}(2014){Mikul{\'a}{\v{s}}ek},
  {Chrastina}, {Li{\v{s}}ka}, {Zejda}, {Jan{\'\i}k}, {Zhu}, \&
  {Qian}}]{Mikulasek2014}
{Mikul{\'a}{\v{s}}ek}, Z., {Chrastina}, M., {Li{\v{s}}ka}, J., {et~al.} 2014,
  Contributions of the Astronomical Observatory Skalnate Pleso, 43, 382

\bibitem[{{Milone} {et~al.}(2010){Milone}, {Kurpi{\'n}ska-Winiarska}, \&
  {Oblak}}]{Milone2010}
{Milone}, E.~F., {Kurpi{\'n}ska-Winiarska}, M., \& {Oblak}, E. 2010, \aj, 140,
  129

\bibitem[{{Moffat}(1986)}]{Moffat1986}
{Moffat}, J.~W. 1986, Canadian Journal of Physics, 64, 178

\bibitem[{{Naoz}(2016)}]{Naoz2016}
{Naoz}, S. 2016, \araa, 54, 441

\bibitem[{{Nelson}(2003)}]{Nelson2003}
{Nelson}, R.~H. 2003, Information Bulletin on Variable Stars, 5371, 1

\bibitem[{{Paxton} {et~al.}(2011){Paxton}, {Bildsten}, {Dotter}, {Herwig},
  {Lesaffre}, \& {Timmes}}]{Paxton2011}
{Paxton}, B., {Bildsten}, L., {Dotter}, A., {et~al.} 2011, \apjs, 192, 3

\bibitem[{{Paxton} {et~al.}(2013){Paxton}, {Cantiello}, {Arras}, {Bildsten},
  {Brown}, {Dotter}, {Mankovich}, {Montgomery}, {Stello}, {Timmes}, \&
  {Townsend}}]{Paxton2013}
{Paxton}, B., {Cantiello}, M., {Arras}, P., {et~al.} 2013, \apjs, 208, 4

\bibitem[{{Paxton} {et~al.}(2015){Paxton}, {Marchant}, {Schwab}, {Bauer},
  {Bildsten}, {Cantiello}, {Dessart}, {Farmer}, {Hu}, {Langer}, {Townsend},
  {Townsley}, \& {Timmes}}]{Paxton2015}
{Paxton}, B., {Marchant}, P., {Schwab}, J., {et~al.} 2015, \apjs, 220, 15

\bibitem[{{Pols} {et~al.}(1997){Pols}, {Tout}, {Schroder}, {Eggleton}, \&
  {Manners}}]{Pols1997}
{Pols}, O.~R., {Tout}, C.~A., {Schroder}, K.-P., {Eggleton}, P.~P., \&
  {Manners}, J. 1997, \mnras, 289, 869

\bibitem[{{Rauw} {et~al.}(2016){Rauw}, {Rosu}, {Noels}, {Mahy}, {Schmitt},
  {Godart}, {Dupret}, \& {Gosset}}]{Rauw2016}
{Rauw}, G., {Rosu}, S., {Noels}, A., {et~al.} 2016, \aap, 594, A33

\bibitem[{{Ribas} {et~al.}(2000){Ribas}, {Jordi}, \& {Gim{\'e}nez}}]{Ribas2000}
{Ribas}, I., {Jordi}, C., \& {Gim{\'e}nez}, {\'A}. 2000, \mnras, 318, L55

\bibitem[{{Ricker} {et~al.}(2015){Ricker}, {Winn}, {Vanderspek}, {Latham},
  {Bakos}, {Bean}, {Berta-Thompson}, {Brown}, {Buchhave}, {Butler}, {Butler},
  {Chaplin}, {Charbonneau}, {Christensen-Dalsgaard}, {Clampin}, {Deming},
  {Doty}, {De Lee}, {Dressing}, {Dunham}, {Endl}, {Fressin}, {Ge}, {Henning},
  {Holman}, {Howard}, {Ida}, {Jenkins}, {Jernigan}, {Johnson}, {Kaltenegger},
  {Kawai}, {Kjeldsen}, {Laughlin}, {Levine}, {Lin}, {Lissauer}, {MacQueen},
  {Marcy}, {McCullough}, {Morton}, {Narita}, {Paegert}, {Palle}, {Pepe},
  {Pepper}, {Quirrenbach}, {Rinehart}, {Sasselov}, {Sato}, {Seager},
  {Sozzetti}, {Stassun}, {Sullivan}, {Szentgyorgyi}, {Torres}, {Udry}, \&
  {Villasenor}}]{Ricker2015}
{Ricker}, G.~R., {Winn}, J.~N., {Vanderspek}, R., {et~al.} 2015, Journal of
  Astronomical Telescopes, Instruments, and Systems, 1, 014003

\bibitem[{{Rosu} {et~al.}(2020){Rosu}, {Noels}, {Dupret}, {Rauw}, {Farnir}, \&
  {Ekstr{\"o}m}}]{Rosu2020}
{Rosu}, S., {Noels}, A., {Dupret}, M.~A., {et~al.} 2020, \aap, 642, A221

\bibitem[{{Sandberg Lacy} \& {Fekel}(2014)}]{Lacy2014}
{Sandberg Lacy}, C.~H., \& {Fekel}, F.~C. 2014, \aj, 148, 71

\bibitem[{{Sandberg Lacy} {et~al.}(1995){Sandberg Lacy}, {Zakirov},
  {Arzumanyants}, {Ishankulov}, {Kharchenko}, {Ibanoglu}, {Tunca}, {Evren},
  {Akan}, \& {Keskin}}]{Lacy1995}
{Sandberg Lacy}, C.~H., {Zakirov}, M., {Arzumanyants}, G., {et~al.} 1995, IBVS,
  4194, 1

\bibitem[{{Schmitt} {et~al.}(2016){Schmitt}, {Schr{\"o}der}, {Rauw},
  {Hempelmann}, {Mittag}, {Gonz{\'a}lez-P{\'e}rez}, {Czesla}, {Wolter}, \&
  {Jack}}]{Schmitt2016}
{Schmitt}, J.~H.~M.~M., {Schr{\"o}der}, K.~P., {Rauw}, G., {et~al.} 2016, \aap,
  586, A104

\bibitem[{{Shakura}(1985)}]{Shakura1985}
{Shakura}, N.~I. 1985, Soviet Astronomy Letters, 11, 224

\bibitem[{{Shao} {et~al.}(2013){Shao}, {Caballero}, {Kramer}, {Wex},
  {Champion}, \& {Jessner}}]{Shao2013}
{Shao}, L., {Caballero}, R.~N., {Kramer}, M., {et~al.} 2013, Classical and
  Quantum Gravity, 30, 165019

\bibitem[{{Smith} \& {Caton}(2007)}]{Smith2007}
{Smith}, A.~B., \& {Caton}, D.~B. 2007, IBVS, 5745, 1

\bibitem[{{Southworth}(2015)}]{Southworth2015}
{Southworth}, J. 2015, in Astronomical Society of the Pacific Conference
  Series, Vol. 496, Living Together: Planets, Host Stars and Binaries, ed.
  S.~M. {Rucinski}, G.~{Torres}, \& M.~{Zejda}, 164

\bibitem[{{Sowell} {et~al.}(2012){Sowell}, {Henry}, \& {Fekel}}]{Sowell2012}
{Sowell}, J.~R., {Henry}, G.~W., \& {Fekel}, F.~C. 2012, \aj, 143, 5

\bibitem[{{Sterne}(1939)}]{Sterne1939}
{Sterne}, T.~E. 1939, \mnras, 99, 662

\bibitem[{{Tkachenko} {et~al.}(2020){Tkachenko}, {Pavlovski}, {Johnston},
  {Pedersen}, {Michielsen}, {Bowman}, {Southworth}, {Tsymbal}, \&
  {Aerts}}]{Tkachenko2020}
{Tkachenko}, A., {Pavlovski}, K., {Johnston}, C., {et~al.} 2020, \aap, 637, A60

\bibitem[{{Tomkin} \& {Fekel}(2006)}]{Tomkin2006}
{Tomkin}, J., \& {Fekel}, F.~C. 2006, \aj, 131, 2652

\bibitem[{{Torres} {et~al.}(2010){Torres}, {Andersen}, \&
  {Gim{\'e}nez}}]{Torres2010}
{Torres}, G., {Andersen}, J., \& {Gim{\'e}nez}, A. 2010, \aapr, 18, 67

\bibitem[{{Torres} {et~al.}(2000){Torres}, {Lacy}, {Claret}, \&
  {Sabby}}]{Torres2000}
{Torres}, G., {Lacy}, C. H.~S., {Claret}, A., \& {Sabby}, J.~A. 2000, \aj, 120,
  3226

\bibitem[{{Torres} {et~al.}(2017){Torres}, {McGruder}, {Siverd}, {Rodriguez},
  {Pepper}, {Stevens}, {Stassun}, {Lund}, \& {James}}]{Torres2017}
{Torres}, G., {McGruder}, C.~D., {Siverd}, R.~J., {et~al.} 2017, \apj, 836, 177

\bibitem[{{Torres} \& {Ribas}(2002)}]{Torres2002}
{Torres}, G., \& {Ribas}, I. 2002, \apj, 567, 1140

\bibitem[{{Torres} {et~al.}(2009){Torres}, {Sandberg Lacy}, \&
  {Claret}}]{Torres2009}
{Torres}, G., {Sandberg Lacy}, C.~H., \& {Claret}, A. 2009, \aj, 138, 1622

\bibitem[{{Torres} {et~al.}(2014){Torres}, {Sandberg Lacy}, {Pavlovski},
  {Feiden}, {Sabby}, {Bruntt}, \& {Viggo Clausen}}]{Torres2014}
{Torres}, G., {Sandberg Lacy}, C.~H., {Pavlovski}, K., {et~al.} 2014, \apj,
  797, 31

\bibitem[{{Torres} {et~al.}(2015){Torres}, {Sandberg Lacy}, {Pavlovski},
  {Fekel}, \& {Muterspaugh}}]{Torres2015}
{Torres}, G., {Sandberg Lacy}, C.~H., {Pavlovski}, K., {Fekel}, F.~C., \&
  {Muterspaugh}, M.~W. 2015, \aj, 150, 154

\bibitem[{{Verma} {et~al.}(2014){Verma}, {Fienga}, {Laskar}, {Manche}, \&
  {Gastineau}}]{Verma2014}
{Verma}, A.~K., {Fienga}, A., {Laskar}, J., {Manche}, H., \& {Gastineau}, M.
  2014, \aap, 561, A115

\bibitem[{{Volkov} \& {Khaliullin}(1999)}]{Volkov1999}
{Volkov}, I.~M., \& {Khaliullin}, K.~F. 1999, IBVS, 4680, 1

\bibitem[{{Will}(2014)}]{Will2014}
{Will}, C.~M. 2014, Living Reviews in Relativity, 17, 4

\bibitem[{{Wolf} {et~al.}(2010){Wolf}, {Claret}, {Kotkov{\'a}},
  {Ku{\v{c}}{\'a}kov{\'a}}, {Koci{\'a}n}, {Br{\'a}t}, {Svoboda}, \&
  {{\v{S}}melcer}}]{Wolf2010}
{Wolf}, M., {Claret}, A., {Kotkov{\'a}}, L., {et~al.} 2010, \aap, 509, A18

\bibitem[{{Wolf} {et~al.}(2006){Wolf}, {Ku{\v{c}}{\'a}kov{\'a}}, {Kolasa},
  {{\v{S}}tastn{\'y}}, {Bozkurt}, {Harmanec}, {Zejda}, {Br{\'a}t}, \&
  {Hornoch}}]{Wolf2006}
{Wolf}, M., {Ku{\v{c}}{\'a}kov{\'a}}, H., {Kolasa}, M., {et~al.} 2006, \aap,
  456, 1077

\bibitem[{{Zasche} \& {Wolf}(2019)}]{Zasche2019}
{Zasche}, P., \& {Wolf}, M. 2019, \aj, 157, 87

\end{thebibliography}

\end{document}